\documentclass[12pt]{article}

\usepackage{amsmath,amssymb,epsfig,bbm,color}
\usepackage{epsfig}
\usepackage{euscript}

\newcommand{\Kcal}{{\cal K}}

\newcommand{\Peu}{\EuScript{P}}
\newcommand{\Keu}{\EuScript{K}}

\newcommand{\Pmf}{\mathfrak{P}}

\newcommand{\Pbbm}{\mathbbm{P}}

\newcommand{\tB}{{\bar{t}}}
\newcommand{\tBl}{{\bar{t}_\lambda}}

\begin{document}

\begin{titlepage}


\begin{flushright}
\bf IFJPAN-IV-2007-3\\
    CERN-PH-TH/2007-059
\end{flushright}

\vspace{1mm}
\begin{center}
  {\LARGE\bf%
    Constrained MC for QCD evolution\\
    with rapidity ordering and minimum kT$^{\star}$\strut
}
\end{center}
\vspace{2mm}

\begin{center}
{\large\bf S.\ Jadach$^{ac}$, W.\ P\l{}aczek$^b$, M.\ Skrzypek$^{ac}$},
\\
{\large\bf P.\ Stephens$^a$}
{\rm and} {\large\bf Z.\ W\c{a}s$^{ac}$}

\vspace{2mm}
{\em $^a$Institute of Nuclear Physics, Polish Academy of Sciences,\\
  ul.\ Radzikowskiego 152, 31-342 Cracow, Poland.}\\ \vspace{1mm}
{\em $^b$Marian Smoluchowski Institute of Physics, Jagiellonian University,\\
   ul.\ Reymonta 4, 30-059 Cracow, Poland.}\\ \vspace{1mm}
{\em $^c$CERN, PH Department, TH Division, CH-1211 Geneva 23, Switzerland.} 
\end{center}

\vspace{3mm}
\begin{abstract}
With the imminent start of LHC experiments, development of phenomenological 
tools, and in particular the Monte Carlo programs and algorithms, becomes urgent.
A new algorithm for the generation of a parton shower initiated 
by the single initial hadron beam is presented.
The new algorithm is of the class of the so called ``constrained MC'' type algorithm 
(an alternative to the backward evolution MC algorithm),
in which the energy and the type of the parton at the end of the parton shower
are constrained (predefined).
The complete kinematics configurations with explicitly constructed
four momenta are generated and tested.
Evolution time is identical with rapidity and minimum transverse momentum
is used as an infrared cut-off.
All terms of the leading-logarithmic approximation 
in the DGLAP evolution are properly accounted for.
In addition, the essential improvements towards the so-called CCFM/BFKL models
are also properly implemented.
The resulting parton distributions are cross-checked up to the $10^{-3}$ 
precision level with the help 
of a multitude of comparisons with other MC and non-MC programs.
We regard these tests as an important asset to be exploited 
at the time when the presented MC will enter as a building block
in a larger MC program for $W/Z$ production process at LHC.
\end{abstract}

\vspace{2mm}
\begin{center}
\em Submitted to Computer Physics Communications
\end{center}

\vspace{3mm}
\begin{flushleft}
{\bf IFJPAN-IV-2007-3\\
     CERN-PH-TH/2007-059\\
     March~2007}
\end{flushleft}

\vspace{2mm}
\footnoterule
\noindent
{\footnotesize
$^{\star}$This work is partly supported by the EU grant MTKD-CT-2004-510126
 in partnership with the CERN Physics Department and by the Polish Ministry
 of Scientific Research and Information Technology grant No 620/E-77/6.PR
 UE/DIE 188/2005-2008.
}

\end{titlepage}

\section{Introduction}

In the past, as at present, the central goal of high energy physics is to 
explore new ranges of energies of interactions. 
These new ranges of energies either facilitate the discovery of new
particles and interactions or validate 
the Standard Model of elementary  interactions, as understood today,
by extending them to an even broader range of energies (distances)
than currently available.
The new generation  of experiments at the nearly completed Large Hadron Collider (LHC)
in Geneva will be ready soon to take data.

For the proper interpretation of expected new data, an 
effort in understanding known physics is needed.
In particular, it might be that the signatures
of the new physics will have to be deciphered from the background
of dominant processes expected from the Standard Model.

In the case of hadron colliders, description of the Standard Model processes 
is rather complicated;
the  colliding hadron beams are not the elementary fields of the
Standard Model  but bounded states of quarks and gluons.
Even worse, in the low energy limit quantum Chromodynamics (QCD) looses
predictive power and does not control the relations between
the hadron wave function and elementary fields representing quarks and gluons.
It is necessary, albeit highly nontrivial, to combine a phenomenological description
of low energy strong interaction phenomena
with the rigorous perturbative QCD predictions at high energies.

A multitude of techniques have been developed to merge low energy aspects 
of strong interaction with the high energy calculations from perturbative QCD.
In this work we concentrate on the methodology based on the so-called
parton distribution functions (PDF) and parton shower Monte Carlo (PSMC).
Special attention will be payed to technical aspects, in particular
to precision testing of the numerical tools.
We believe that this is very important for the future efforts
in minimizing overall systematic errors of the QCD prediction
in which PSMCs are used.

In this work we shall concentrate on the question of the
evolution equation of the PDF in the Monte Carlo (MC) form.
Such an evolution equation for the initial state hadron describes how 
the PDF responds to an increase of
the dimensional, large energy scale $Q=\mu$,
set by the hard process probing the PDF.
The formula for the integrated cross section,
which combines the hard process with the matrix element, has been proved
within perturbative QCD in a form of the so called 
{\em factorization theorems},
see for instance refs. \cite{Collins:1984kg,Bodwin:1984hc}.
These theorems have been proven starting directly from the Feynman diagrams,
integrated over the phase space
and convoluted with the nonperturbative parton wave function in a hadron.
The evolution equation of the PDF can also be formulated using renormalization
group and operator product expansion \cite{stirling-book}.

However, for the real-life practice of the present and future hardon-hadron
and hadron-electron collider experiments, one needs a more refined (exclusive)
picture of the multiparton production, the simplest one
being the so called parton shower, governed by perturbative QCD.
In principle, it should reproduce the evolution of PDFs,
after integrating it over the Lorentz invariant phase space.
If the above is conveniently implemented in the form of the parton shower
Monte Carlo event generator,
see for example \cite{Sjostrand:2000wi,Corcella:2000bw},
with the inclusion of the parton hadronization process --
a very useful feature for the collider experiments.
In the construction of such a PSMC the evolution equation of
the PDF is used as a guide for defining distributions of the partons emitted
from a single energetic hadron (the initiatial parton in the shower)%
\footnote{This is clearly a kind of ''backward engineering'' --
  it would be better to get distributions of partons forming
  PDF at large $Q$ directly from the Feyman diagrams. 
  Unfortunately it is too difficult.}.

Having in mind the above context, this work has several aims.
The principal aim is to use once again the evolution equation of the PDF
in order to model the multiparton parton shower initiated by the single
parton located inside the single hadron of the collider beam.
We shall insist that, as in the factorization theorems,
this modelling has to be {\em universal},
that is independent from showering of the other hadron beam,
spectator partons, and the type of the {\em hard process} of
the parton-parton scattering at the large energy scale $Q$.

Another aim is to define and maintain a clear prescription relating variables in the
evolution of the PDF with the four-momenta in the PSMC.
We keep in mind that such a PSMC will be a {\em building block} to be used
for two beams%
\footnote{This will be done in the forthcomming paper
  on the new parton shower MC for $W/Z$ production
  in hadron collider of the forthcomming paper \cite{DoubleCMC}.},
hence the upper limit of the multiparton phase
space (related to $Q$) should allow for smooth coverage of the
entire phase space, without any gaps and overlaps.

There is also an important technical problem to be addressed:
the off-shell parton entering the hard process has to have
predefined energy and flavour matching preferences of the hard
process; hence, a Markovian MC 
(which is a natural MC implementation of the PSMC)
cannot be used to model the initial state PSMC.
However, instead of using the so called 
{\em backward evolution} \cite{Sjostrand:1985xi},
our choice will be to employ the technique of the {\em constrained} MC
(referred to as CMC technique);
that is to generate the multiparton distribution with the
restriction on the value of the parton energy and type of the parton.
Two distinct versions of this relatively new technique
were proposed and tested in
refs.~\cite{Jadach:2005bf,Jadach:2005yq,Jadach:2005rd}
for the DGLAP~\cite{DGLAP} evolution in the leading-logarithmic (LL)
approximation.

The aim of this work is to extend the most promising variant 
of the above CMC technique to
a wider class of evolution kernels, beyond DGLAP, 
towards evolution models of the CCFM class~\cite{CCFM};
maintaining at the same time the
explicit mapping of the evolution variables into four-momenta.

The other important longer term goal is to facilitate
the inclusion of the complete NLO
corrections by means of the clearer/cleaner modelling of the PSMC,
as compared with the existing combined NLO and PSMC 
calculations~\cite{Frixione:2002ik,Frixione:2006he}.
This will be achieved, for example,
by means of better coverage of the phase space
in the basic parton shower MC.

The outline of the paper is the following: 
In section 2 we shall formulate the general formalism of the evolution equations
and their solutions in a form suitable for the CMC technique.
In section 3 we discuss two particular types of the kernels
and related Sudakov form-factors.
In section 4 we outline the CMC algorithm for the pure gluonstrahlung segments.
In section 5 we shall introduce in the CMC quark gluon transitions.
In section 6 results of precision numerical tests of our CMC implementation
will be reported.
For testing CMC implementations we shall use auxiliary Markovian MC programs
which are described and tested in separate papers \cite{SingleMMC}.
Finally, we summarize the main results; some technical details will
be included in the appendices.

\section{Evolution equations and solutions}
In the following we shall formulate the mathematical framework
for the evolution equation and its solutions in a form suitable
for the construction of the CMC algorithm in the latter part of the paper.

The generic evolution equation covering several types of evolution reads
\begin{equation}
\label{eq:genevoleq}
  \partial_t D_f(t,x)
 = \sum_{f'} \int_x^1 du\; \Keu_{ff'}(t,x,u) D_{f'}(t,u).
\end{equation}
It describes the evolution of the parton distribution function $D_j(t,u)$,
where $x$ is fraction of the hadron momentum carried by the parton
and $j$ is the type (flavour) of the parton.
The variable $t=\ln Q$ is traditionally
called an {\em evolution time} and it represents the (large)
energy scale $Q=\mu$ at which the PDF is probed using a hard scattering process.
The LL DGLAP case \cite{DGLAP} is covered by eq.~(\ref{eq:genevoleq})
with the following identification
\begin{equation}
\label{eq:dglap-kernel}
  \Keu_{ff'}(t,x,u) = \frac{1}{u}\;{\Peu}_{ff'}\left(t,\frac{x}{u}\right)
  = \frac{\alpha_S(t)}{2\pi} \frac{2}{u}\; P^{(0)}_{ff'}\left(t,\frac{x}{u}\right),
\end{equation}
where $P^{(0)}_{ff'}(z)$ is the standard LL DGLAP kernel and the
factor 2 is related to our definition of the evolution variable $t$. 

In the compact operator (matrix) notation eq.~(\ref{eq:genevoleq}) reads
\begin{equation}
\label{eq:op-evoleq}
  \partial_t {\bf D}(t) = {\bf K}(t) \; {\bf D}(t).
\end{equation}
Given a known ${\bf D}(t_0)$, the formal solution at any later ``time'' $t\geq t_0$
is provided by the time ordered exponential
\begin{equation}
\label{eq:timex} 
  {\bf D}(t) 
   = \exp\left( \int_{t_0}^t {\bf K}(t') dt' \right)_T \; {\bf D}(t_0)
   = {\bf G}_{\bf K}(t,t_0)\; {\bf D}(t_0).
\end{equation}
The {\em time-ordered exponential} evolution operator reads%
\footnote{Here and in the following we define 
$\prod_{i=1}^n A_i \equiv A_n A_{n-1}\dots A_2A_1$.}
\begin{equation}
\label{eq:torder}
{\bf G}_{\bf K}(t,t_0)=
{\bf G}({\bf K};t,t_0)=
\exp\left( \int_{t_0}^t {\bf K}(t') dt' \right)_T
 = {\bf I} +\sum_{n=1}^\infty 
      \prod_{i=1}^n  \int_{t_0}^t dt_i \theta_{t_i>t_{i-1}} {\bf K}(t_i),
\end{equation}
For the sake of completeness, let us write the explicit definition%
\footnote{In the case of QCD evolution ${\bf K}(t_i)$ transforms $u\in(0,1)$
  into $x$ obeying $x<u$. This is due to 4-momentum conservation.
}
of  the multiplication operation as used and  defined in 
eqs.~(\ref{eq:op-evoleq}-\ref{eq:torder}):
\begin{equation}
\label{eq:clarity}
   \big( {\bf K}(t_2)\;{\bf K}(t_1)\big)_{f_2,f_1}(x_2,x_1)=
   \sum_{f'} \int_{x_2}^{x_1} dx'\;
        \Keu_{f_2f'}(t_2,x_2,x')\; \Keu_{f'f_1}(t_1,x',x_1).
\end{equation}
We shall often be dealing with the case of the kernel split into parts,
for example:
\begin{equation}
     {\bf K}(t)={\bf K}^A(t)+{\bf K}^B(t).
\end{equation}
In such a case the solution of eq.~(\ref{eq:timex}) can be reorganized
as follows%
\footnote{ Here and in the following,
  we understand that the scope of the indices ceases at
  the closing bracket. However, the validity scope of indiced variables,
  for instance of $t_i$, extends until the formula's end.
  The use of eq.~(\ref{eq:clarity}) is understood to be adjusted
  accordingly.
}
\begin{equation}
\label{eq:timex2}
\begin{split}
{\bf D}(t)
&={\bf G}_{{\bf K}^B}(t,t_0)\; {\bf D}(t_0)
+\sum_{n=1}^\infty
 \left[
 \prod_{i=1}^n \int_{t_0}^{t}dt_i\; \theta_{t_i>t_{i-1}}
    {\bf G}_{{\bf K}^B}(t_{i+1},t_{i})\;
    {\bf K}^A(t_{i})
 \right]
 {\bf G}_{{\bf K}^B}(t_{1},t_{0})\; {\bf D}(t_0)
\\&
={\bf G}_{{\bf K}^B}(t,t_0)\; {\bf D}(t_0)
+\sum_{n=1}^\infty
 \left[\prod_{i=1}^n \int_{t_{i-1}}^{t}dt_i\right]
 {\bf G}_{{\bf K}^B}(t,t_{n})\; 
 \left[\prod_{i=1}^n
    {\bf K}^A(t_{i})\;
    {\bf G}_{{\bf K}^B}(t_{i},t_{i-1})
 \right]
{\bf D}(t_0),
\end{split}
\end{equation}
where $t_{n+1}\equiv t$ and ${\bf G}_{{\bf K}^B}$ is the evolution operator
of eq.~(\ref{eq:torder}) of the evolution with the kernel ${\bf K}^B$.
Formal proof of eq.~(\ref{eq:timex2}) is given in  ref.~\cite{Jadach:2006yr}.

\begin{figure}[!ht]
  \centering
  {\epsfig{file=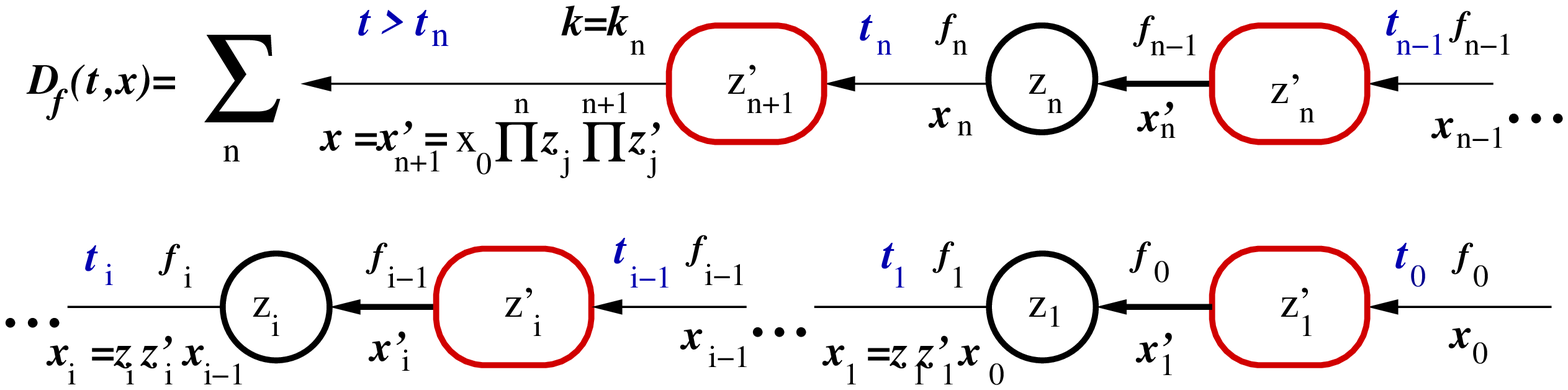,width=140mm}}
  \caption{\sf
    The scheme of integration variables and summation indices 
    in eq.~(\protect\ref{eq:SoluX}); the circles correspond to ${\bf K}^A(t_{i})$ 
    and the ovals to
    $ {\bf G}_{{\bf K}^B}$, $z_i'=x_i'/x_{i-1}$ and $z_i=x_i/x_{i}'$.
    }
  \label{fig:hierarchX}
\end{figure}

In the following we are going to nest eq.~(\ref{eq:timex2}) twice.
First, we employ it in order to isolate gluonstrahlung and flavour-changing
parts of the evolution, exploiting the following split of the kernel
\begin{equation}
     \Keu(t)_{ff'}= \Keu^A(t)_{ff'}+\Keu^B(t)_{ff'}
                 = (1-\delta_{ff'})\Keu(t)_{ff'}+\delta_{ff'}\Keu(t)_{ff}.
\end{equation}
In this case ${\bf G}_{{\bf K}^B}$ represents pure gluonstrahlung
and is diagonal in the flavour index.
In the standard integro-tensorial notation eq.~(\ref{eq:timex2}) looks as follows:
\begin{equation}
\label{eq:SoluX}
\begin{split}
D_f(t,x) =&
     \int\limits_x^1 dx_0\; 
     G_{ff}(K^B;t, t_{0};x,x_0)\;
     D_f(t_0,x_0)
  +\sum_{n=1}^\infty \;
   \sum_{f_{n-1},\dots,f_{1},f_{0}}
\\ \times&
    \int\limits_{x}^1 dx_0\;
   \Bigg[ \prod_{i=1}^n \int\limits_{t_{i-1}}^t dt_i\;
     \int\limits_0^1 dx_i\;
     \int\limits_0^1 dx'_i\;
     \theta_{x_i<x'_i<x_{i-1}}
     \Bigg]\;
   G_{ff}(K^B;t,t_n,x,x_n)\;
\\ \times&
   \Bigg[ \prod_{i=1}^n
     \Keu^A_{f_if_{i-1}}(t_i,x_i,x'_i)\;
     G_{f_{i-1}f_{i-1}}(K^B;t_i,t_{i-1},x'_i,x_{i-1}) \Bigg]\;
      D_{f_0}(t_0,x_0),
\end{split}
\end{equation}
where $f_n\equiv f$.
In the above and the following equations we adopt the following notation:
\[
  \delta_{x=y}=\delta(x-y)
\]
and
\[
  \theta_{y<x}=1~~~\hbox{\rm for}~~~y<x~~~\hbox{\rm and}~~~
  \theta_{y<x}=0~~~\hbox{\rm for}~~~y\geq x.
\]
Similarly, $\theta_{z<y<x}=\theta_{z<y}\theta_{y<x}$.
The chain of integration variables and flavour indices is depicted 
schematically in Fig.~\ref{fig:hierarchX}.

Next, eq.~(\ref{eq:timex2}) is used in order to resum the virtual
IR-divergent part $\Keu^V$ of the gluonstrahlung kernel $\Keu^B$
\begin{equation}
\begin{split}
&
 \Keu^B_{ff'}(t,x,u)= \delta_{ff'}\Keu_{ff}(t,x,u)
     = \delta_{ff'}\;\big(\Keu^V_{ff}(t,x,u) +\Keu^R_{ff}(t,x,u)\big),
\\&
 \Keu^V_{ff}(t,x,u)= -\delta_{x=u}\Keu^v_{ff}(t,x),
\\&
 \Keu^R_{ff}(t,x,u)= \theta_{x<u-\Delta(x,u)} \Keu_{ff}(t,x,u),
\end{split}
\end{equation}
where $\Delta(x,u)$ is finite IR cut-off, not necessarily
infinitesimal. 
In order to resum (exponentiate) the virtual part $\Keu^V$ of the kernel
the following version of eq.~(\ref{eq:timex2}) is employed
\begin{equation}
\begin{split}
&{\bf G}_{{\bf K}^V+{\bf K}^R}(t,t_0)=
 {\bf G}_{{\bf K}^V}(t,t_0)\; +
\\&~~~~~~~~~~~~~
+\sum_{n=1}^\infty
 \left[
 \prod_{i=1}^n \int_{t_0}^{t}dt_i\; \theta_{t_i>t_{i-1}}
    {\bf G}_{{\bf K}^V}(t_{i+1},t_{i})
    {\bf K}^R(t_{i})
 \right]
 {\bf G}_{{\bf K}^V}(t_{1},t_{0}).
\end{split}
\end{equation}
Since  ${\bf K}^V(t_{i})$ is diagonal in $x$, $u$ and in the flavour index,
and also because of
\begin{equation}
\{G_{{\bf K}^V}(t_{i+1},t_{i})\}_{ff}(x,u)
  =\delta_{x=u}\; e^{-\Phi_f(t_{i+1},t_{i}|x)},\qquad
\Phi_f(t_{i+1},t_{i}|x)=\int_{t_{i}}^{t_{i+1}} dt\; \Keu^v_{ff}(t,x),
\end{equation}
we obtain immediately
\begin{equation}
\label{eq:EvolsoluG}
\begin{split}
G_{ff}(K^B;t_{b},t_{a},x,u) 
  \equiv \{& {\bf G}_{{\bf K}^B}(t_{b},t_{a})\}_{ff}(x,u) =
\\
  = e^{-\Phi_f(t_b,t_a|x)} \delta_{x=u}
  +&\sum_{n=1}^\infty \;
      \bigg[ \prod_{i=1}^n 
      \int_{t_a}^{t_b} dt_i\; \theta_{t_i>t_{i-1}}  
      \int_x^u dx_i
     \theta_{x_i<x_{i-1}}
      \bigg]
\\&\times
   e^{-\Phi_f(t_b,t_n|x)}
   \bigg[\prod_{i=1}^n 
        \Keu^R_{ff} (t_i,x_i,x_{i-1}) 
         e^{-\Phi_{f}(t_i,t_{i-1}|x_{i-1})} \bigg]
   \delta_{x=x_n},
  \end{split}
\end{equation}
where $t_{n+1}\equiv t_b$, $t_{0}\equiv t_a$ and $x_0\equiv u$.
The above result can also be obtained
by iterating the evolution equation for ${\bf G}_{{\bf K}^B}(t,t_0)$
with the boundary condition ${\bf G}_{{\bf K}^B}(t_0,t_0)={\bf I}$,
see for instance ref.~\cite{Golec-Biernat:2006xw}.

The  algebra resulting from (\ref{eq:timex2}) is quite general
and does not rely on any particular form of the kernel. 
For example, in the above calculations
we did not have to invoke the energy sum rules, or any other specific
restrictions on the overall normalization embodied
usually in the virtual corrections.
Also, we did not define yet the relation between  the evolution
variables $t_i$ and $x_i$  and parton  four-momenta.
This will be done in the following section.

\section{Evolution kernel and variables}
Before we specify details of the evolution kernels used
in this work, let us discuss the relation between the evolution variables
$x_j$, $t_j$ and the emitted parton four-momenta $k_j^\mu$
in the corresponding parton shower MC.

In the proofs of the factorization theorems \cite{Collins:1984kg,Collins:1981uk,Bodwin:1984hc}
and its practical realizations like in ref.~\cite{Curci:1980uw},
typically within $\overline{MS}$ scheme, one projects the four-momenta
of the (off-shell) partons
into the 1-dimensional variable  of the evolution,
typically the dimensionless lightcone variable $x$.
The so-called factorization scale $\mu_F=Q$ measuring the size
of the available parton emission phase space is usually set by the 
kinematics of the hard process.
The underlying QCD differential distribution 
(the QCD matrix element times the phase space) is reduced to
a chain of parton splittings with the $x$ variable of the PDF
being the fraction of the initial hadron energy $E_h$
carried by the parton entering the hard process.
The variable $e^{t}=Q$ defines the boundary (maximum value)
for many ordered variables%
\footnote{ This simplified picture is valid at least 
           in the leading-logarithmic approximation.}
$t_i$.
The variable $t$ may be related directly to one of the phase space parameters
(virtuality, transverse momentum, angle) or the abstract
dimensional scale variable $\mu_F$ resulting from 
the formal procedure of cancelling IR singularities
in the dimensional regularization method.
In the classical construction of the parton shower MC, one must
invert the above mapping of the phase space variables into the
evolution variables; that is to construct parton four-momenta out of
$t_i$ and $x_i$ and to reconstruct fully differential 
parton distributions in terms of these four-momenta.
Obviously this procedure is not unique and requires some guidance from
the detailed knowledge of the structure of the IR singularities%
\footnote{We understand IR singularities as both the collinear
          and soft ones.}
of the original QCD matrix element.

\begin{figure}[!ht]
  \centering
  {\epsfig{file=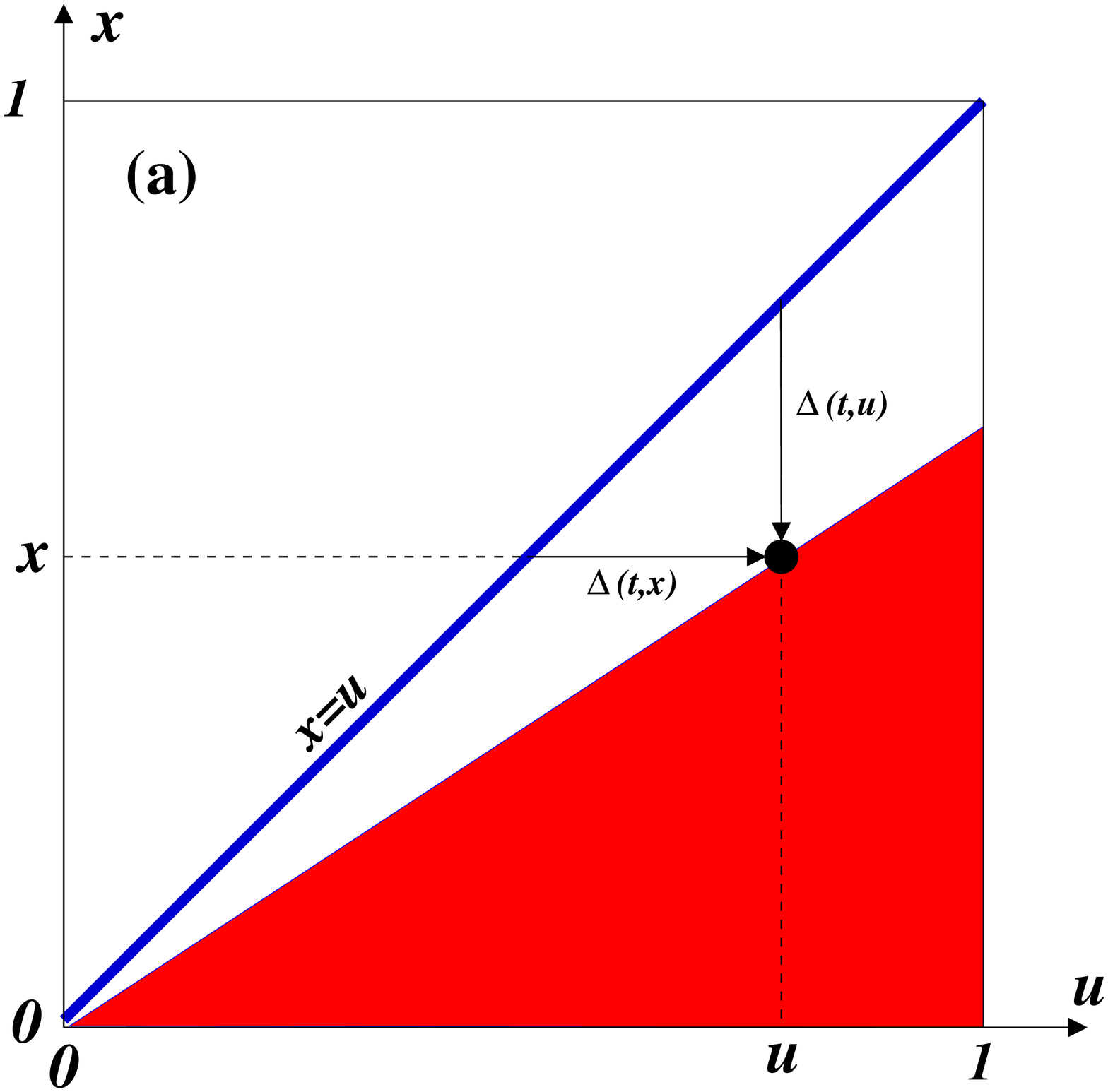,width=70mm}}~~
  {\epsfig{file=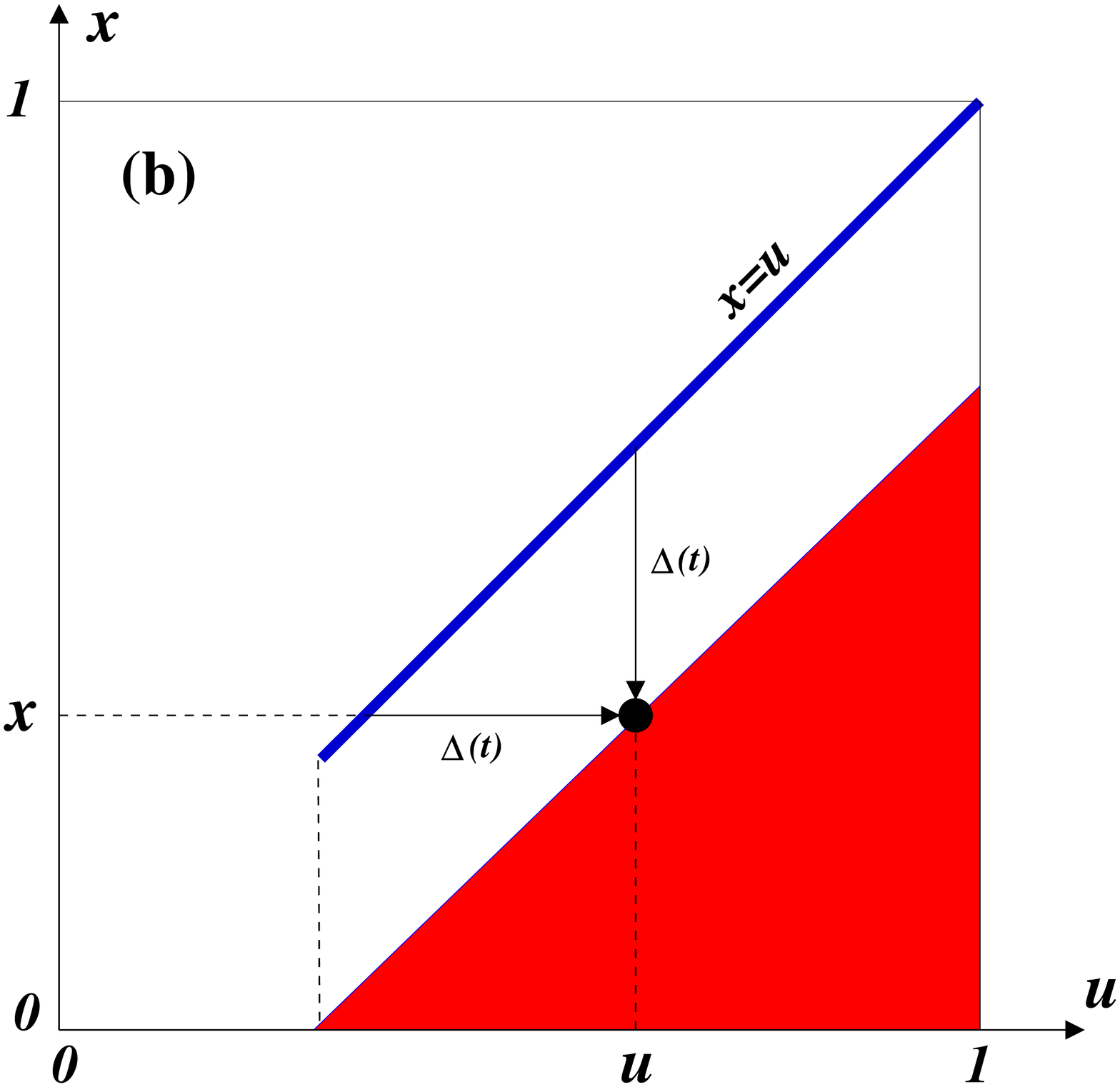,width=70mm}}
  \caption{\sf
    Two types of the infrared (IR) boundaries on the $u-x$ plane
    (a) $u-x>u\epsilon$ and (d) $u-x>\epsilon$ for the
    triangle (red) which depicts the area where the real emission part
    $\Kcal^\theta$ is nonzero. The  
    diagonal line (blue) represents places where
    the virtual $\Kcal^v$ is nonzero.
    }
  \label{fig:x-u-plane}
\end{figure}

In this work we shall construct the CMC algorithm for
two new types of generalized kernels,
in addition to the ones of LL DGLAP of eq.~(\ref{eq:dglap-kernel})
for which examples of CMC were already constructed in
refs.~\cite{Jadach:2005bf} and \cite{Jadach:2005yq}.
The new kernels are based on the following generic form
\begin{equation}
\label{eq:generic-kernel}
\begin{split}
&\Kcal_{ff'} (t, x, u) 
   = \frac{\alpha_S (t, x, u)}{\pi}  \frac{1}{u}
    P_{ff'}^{(0)} \left( t, \frac{x}{u} \right)
    = -\Kcal^v_{ff} (t, u) \delta_{ff'} \delta_{x = u}
    +\Kcal^{\theta}_{ff'} (t, x, u),
\\&
\Kcal^{\theta}_{ff'} (t, x, u)=\Kcal_{ff'} (t, x, u)
            \theta_{u \geq x + \Delta (t,u)},
\end{split}
\end{equation}
where $P_{ff'}^{(0)}(t,z)$ is the standard LL kernel (DGLAP) and $t$-dependence
enters into it only through the IR regulator $\Delta(t,u)$.
The new kernel types correspond to different choices of the argument
of the strong coupling constant and to different forms of the
$\Delta(t,u)$ regulator. 
The two new types of the regulator $\Delta(t,u)$ used in this
work are depicted on the $x-u$ plane in Fig.~\ref{fig:x-u-plane} and
will be defined in the following section.

The other important departure from DGLAP is 
the strong coupling constant $\alpha_S(t,x,u)$ 
which now may  depend
on all evolution variables.
We shall consider the strong coupling constant $\alpha_S$
depending on  $z=x/u$ or on the transverse momentum $k^T$ defined below.
Before we define the evolution kernel in a detail,
we have to elaborate first the mapping
of the evolution variables into four-momenta.

\subsection{Relating evolution variables to four-momenta}
\label{sec:kinema}
The essential decision in the construction of the parton shower MC
concerns the choice
of a kinematics variable in the solutions of the evolution
equations~(\ref{eq:SoluX}) and (\ref{eq:EvolsoluG}); this choice is in one-to-one correspondence with
the evolution time variable $t_i$ and the limiting value $t$.
We choose to associate $t$ with the rapidity (angle)
of the emitted parton, following the well known arguments
on the colour coherence exposed in many papers,
see for instance refs.~\cite{Mueller:1981ex,Catani:1990rr,khoze-book}
and further references therein.

\begin{figure}[!ht]
  \centering
  {\epsfig{file=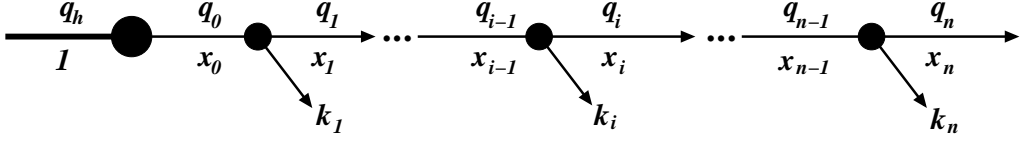,width=140mm}}
  \caption{\sf
    Kinematics in the evolution history (tree).
    }
  \label{fig:kine-evol}
\end{figure}

We define the lightcone variables $q^{\pm} = q^0 \pm q^3$ and 
normalize the parton momenta with respect to the energy $E_h$
of the initial massless hadron, see Fig.~\ref{fig:kine-evol},
\begin{equation}
  q_h^+ = 2 E_h,\quad {\rm and}\quad
  q_i^+ = x_i 2 E_h,\quad {\rm where}\quad
  0 \leq x_i \leq 1.
\end{equation}
These relations hold in the rest frame of the hard process system (HRS)
with the $z$-axis along the momentum of initial state hadron $h$.

In particular, for the parton initiating a parton cascade, we have
\begin{equation}
   q_0^+ = x_0 2 E_h.
\end{equation}
This parton has negligible transverse momentum%
\footnote{In the realistic MC $k^T_0$ will be distributed according 
 to a Gaussian profile with the width of $\sim {\cal O}(\lambda)$.},
$k_0^T = 0$.
In the HRS each on-shell $i$-th {\em emitted particle}
will take away a part of the lightcone variable
\begin{equation}
  k_i^+ = q_{i - 1}^+ - q_i^+ = (x_{i - 1} - x_i) 2 E_h .
\end{equation}
The above is not enough to define $k^{\mu}$.
For this we need to define at least 
$k^T = |\vec{k}^T|$ or the rapidity $\eta_i=\frac{1}{2}\ln (k_i^+/k^-_i)$.
Once the azimuthal angle $\varphi_i$ is added, we can complete  the mapping
from the evolution variables to the 4-momentum of the $i$-th emitted parton
$k^{\mu}_i (t_i, x_i, x_{i - 1},\varphi_i)$.

If we associate the evolution time $t_i$ with the rapidity $\eta_i$
then the following relation
\begin{equation}
  k_i^T = \sqrt{k_i^+ k^-_i} = k_i^+ \sqrt{\frac{k^-_i}{k^+_i}} = k_i^+ e^{-
  \eta_i} = (x_{i - 1} - x_i) 2 E_h e^{- \eta_i},
\end{equation}
valid for $k^2_i = 0$, provides the transverse momentum and thus $k_i^\mu$.
 We can also eliminate $ k_i^T$ 
with the help of  the following conventional relation which defines
the evolution time $t$
\begin{equation}
  k_i^T \equiv e^{t_i} (x_{i-1}-x_i).
\end{equation}
Note that this relation translates into $k^T = e^t (u-x)$ in the
general evolution equation of eq.~(\ref{eq:genevoleq}).
We observe that  evolution time $t_i$ and rapidity $\eta_i$ are related 
by a linear transformation of  the following explicit form:
\begin{equation}
\label{eq:dwojka}
e^{t_i} = e^{\ln (2 E_h) -\eta_i}\quad
\Rightarrow\quad \eta_i = \ln (2 E_h)-t_i.
\end{equation}
The above relation is the main result of this section.

Summarizing, the mapping of $t_i$ and $x_i$ into 4-momenta
$k^{\mu}_i, i = 1,2, \ldots n$,
can be written now in an explicit manner:
\begin{equation}
\begin{split}
&  k_i^+  =  (x_{i - 1} - x_i) 2 E_h,\qquad
  k_i^T  =  (x_{i - 1} - x_i) e^{t_i},\qquad
  k_i^-  =  (k_i^T)^2 / k_i^+,
\\&
  k_i^0  =  (k_i^+ + k_i^-) / 2,\qquad~~~~ k_i^3 = (k_i^+ - k_i^-) / 2.
\end{split}
\end{equation}

Last but not least, we have to define also the phase space limits,
$t_i\in(t_{\min}, t_{\max})$ and $\eta_i \in(\eta_{\min},\eta_{\max})$.
One has to be very careful at this step.
The maximum evolution time $t_{\max}$ 
(minimum rapidity $\eta_{\min}$) is set by
the requirement that all emitted partons are confined to the forward hemisphere%
\footnote{A sharp minimum/maximum rapidity is essential for avoiding  mismatch
  between the parton distributions from two independent
  constrained MCs ``operating'' in the backward
  and forward hemispheres for the initial-state radiation.},
$90^\circ \geq \theta_i$, or equivalently $\eta_i\geq \eta_{\min}=0$.
This implies
\begin{equation}
  t_{\max} = \ln (2 E_h),\quad E_h = \frac{1}{2} e^{t_{\max}}.
\end{equation}

The minimum evolution time is determined by the phase space opening point
for the first emission, due to minimum transverse momentum
$k^T_{\min}=\lambda$:
\begin{equation}
e^{t_1} x_0 \geq \lambda.
\end{equation}
This leads to
\[
   t_1 > t_{_{\lambda}} - \ln x_0,\quad t_\lambda\equiv \ln\lambda
\]
and therefore to
\[ 
   t > t_n > t_{n - 1} > \ldots > t_2 > t_1 > t_{_{\lambda}}-\ln x_0
   \equiv t_{\min}.
\]
This automatically determines the maximum rapidity
\[ \eta_{\max} = \ln (2 E_h) - \ln \lambda +\ln x_0. \]
Altogether 
\begin{equation}
\begin{split}
& t_i  \in (t_{\min}, t_{\max}) = (\ln \lambda-\ln x_0, \ln (2 E_h)),
\\&  
\eta_i \in  (\eta_{\min}, \eta_{\max}) 
       = (0, \ln (2 E_h)+\ln x_0 - \ln \lambda)),
\\&
\eta_{\max} - \eta_{\min} =  t_{\max} -t_{\min} = \ln (2 E_h / \lambda).
\end{split}
\end{equation}

\begin{figure}[!ht]
  \centering
  \epsfig{file=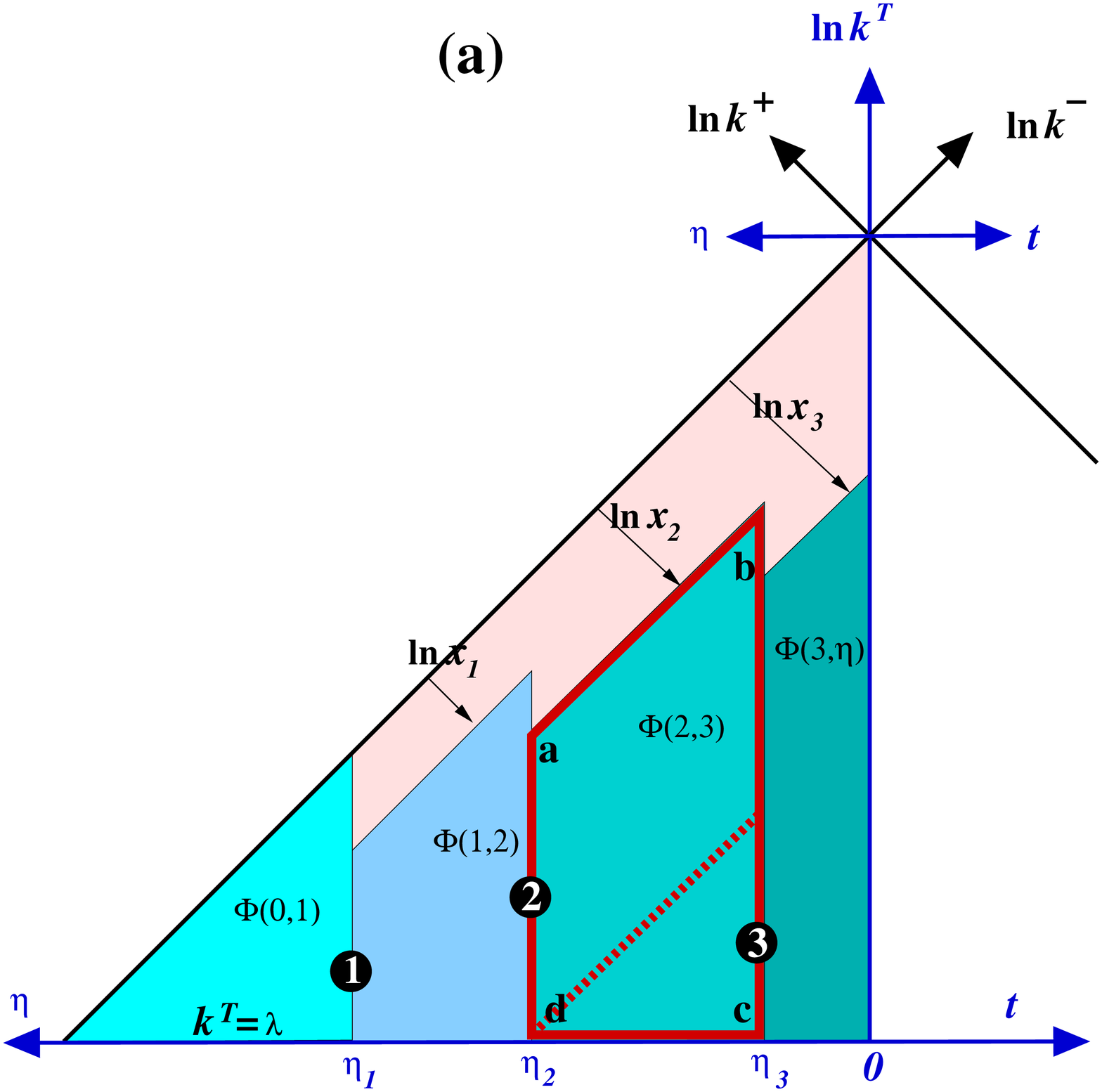, width=75mm}\;
  \epsfig{file=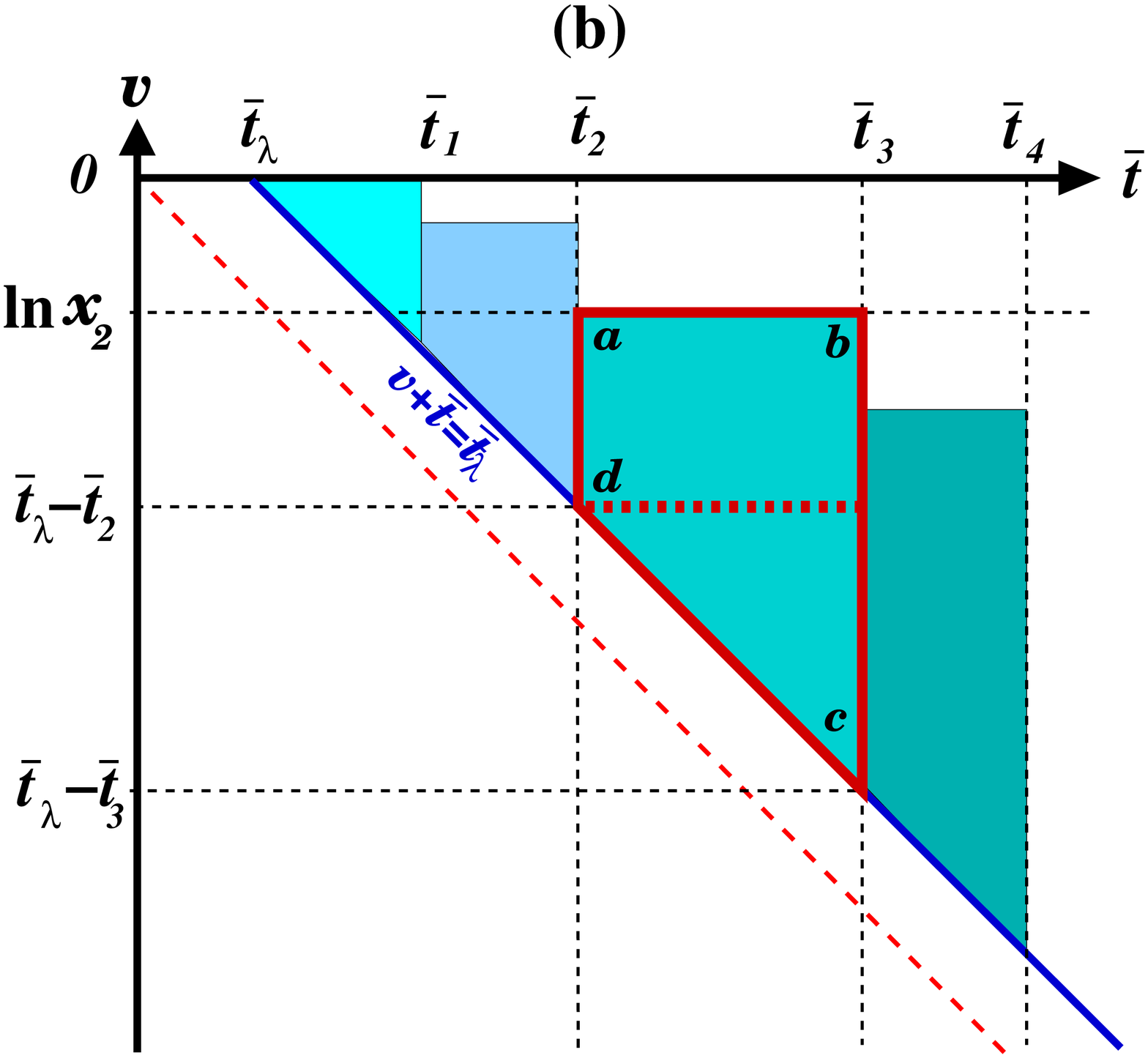, width=75mm}
  \caption{\sf
  Sudakov plane parametrized
  using two different sets of variables:
  (a) $(t,v)$ and/or $(\eta,\ln k^T)$,
  (b) $(\bar{t},v)$, where $\bar{t}=t-\ln\Lambda_0$, see Appendix.
  For simplicity $x_0=1$ is set.
  Dashed red line marks position of the Landau pole.
    }
  \label{fig:sudak0}
\end{figure}

Let us remark that the naive assignment $t_{\max} = \ln (E_h)$,
without the factor of 2, would lead,  because of eq.~(\ref{eq:dwojka}), 
to a partial coverage of the forward hemisphere only,
$\eta_i \geq \ln (2)$.
On the other hand, this factor of 2 may look justified;
the absolute kinematic range of the transverse momentum
is $k_i^T \in (\lambda, k_{\max}^T)$, where $k_{\max}^T = e^{t_{\max}} = 2
E_h$ results from the relation $k_i^T \equiv e^{t_i} (x_{i - 1} - x_i)$
and energy conservation $x_i \leqslant 1$.
The reader may notice
that this limit is a factor of 2 higher than in the familiar
inequality $k^T_i \leq E_h$ resulting from the 4-momentum conservation
operating in both hemispheres simultaneously,
i.e. on $k_i^+$ and $k_i^-$.
However, our limit is valid, including the factor of 2, for a single hemisphere
separated from any ``activity'' on the other side!

Finally, we illustrate the real emission phase space
in Fig.~\ref{fig:sudak0}(a) using the rapidity variable $\eta$ and
the log of transverse momentum $k^T$.
Directions of the lightcone variables $k^\pm$ are also indicated.
In this figure we indicate, as black numbered points, momenta of three
example emitted partons.
Available phase space is limited from below by $k^T_{\min}=\lambda$,
while in the direction $k^+$ the boundary is controlled
by the total available $q^+$, which is
diminished by the factor $x_{i-1}$ at $i$-the step
(setting for simplicity $x_0=1$).
The minimum rapidity $\eta=0$ limits the emission phase space from the right
hand side.
The triangle and trapezoids show integration domains of the consecutive
form-factors $\Phi_f(t_i,t_{i-1}|x_{i-1})$ in eq.~(\ref{eq:EvolsoluG}),
see also section \ref{sec:formfactor} and Appendix~\ref{append:splane}.

The above mapping of the evolution variables
into four-momenta, with the minimum $k^T$ and built-in angular ordering,
is identical to the one used in 
refs.~\cite{Marchesini:1990zy,Marchesini:1991zy,Marchesini:1995wr},
up to the following redefinition of the evolution time:
$\hat{t}_i = \ln E_h - \eta_i + \ln x_{i - 1} = t_i + \ln x_{i - 1}$.
For the redefined $\hat{t}$ we have
\[ \hat{t} + \ln x > \hat{t}_n + \ln x_n > \hat{t}_{n-1} + \ln x_{n-1} > 
  \ldots > \hat{t}_2 + \ln x_2 > \hat{t}_1 + \ln x_1 > t_{_{\lambda}}, \]
and consequently $ \hat{q}_i >\hat{q}_{i-1} z_{i-1}$, with $z_i=x_i/x_{i-1}$.

\subsection{Evolution kernels}
\label{sec:kernels}
Once the phase-space parametrization is explained, we may define our choices 
for the evolution kernels, introduced so far only in the generic form in
eq.~(\ref{eq:generic-kernel}).
The kernels to be used in the CMC in this work will be of three kinds.
The main difference between them is 
in the choice of the variable used as an argument
of the coupling constant $\alpha_S$.
The appearance of the Landau pole in $\alpha_S$ will limit
the choice of the IR cut-off in the multigluon phase space.
Let us define first the gluonstrahlung kernel $f=f'$,
where $f$ is flavour type, $f=G,q,\bar{q}$, in all three cases.
The strong coupling constant
will be always taken in the LL approximation
\begin{equation}
  \alpha_S^{(0)}(q)=\frac{2\pi}{\beta_0}\; \frac{1}{\ln q-\ln\Lambda_0}.
\label{eq:alpha}
\end{equation}

{\em Case (A)}: The standard DGLAP LL 
of ref.~\cite{Jadach:2005bf}, which is used here as a reference case:
\begin{equation}
\label{eq:kernBremsA}
\begin{split}
\Kcal^{\theta(A)}_{ff} (t, x, u)
&=\frac{\alpha_S (e^t)}{\pi}  \frac{1}{u}
    P_{ff}^{(0)}(z)\; \theta_{1-z\geq \epsilon}
 =\frac{\alpha_S (e^t)}{\pi}  \frac{1}{u}
    P_{ff}^{(0)}(x/u)\; \theta_{u-x\geq u\epsilon},
\end{split}
\end{equation}
where $\epsilon$ is infinitesimally small and $z=x/u$.

{\em Case (B)}: The argument in $\alpha_S$ is $(1-z)e^t=k^T/u$;
such a choice  was already advocated in 
the early work of ref.~\cite{Amati:1980ch}.
For the IR cut-off we use $\Delta(t,u)=\lambda ue^{-t}$,
where $\lambda>\Lambda_0$. It cannot be infinitesimally small,
however, it becomes very small at large $t$.
Using $z=x/u$ we define
\begin{equation} 
\label{eq:kernBremsB}
\begin{split}
\Kcal^{\theta(B)}_{ff} (t, x, u)
&=\frac{\alpha_S^{(0)} ((1-z)e^t)}{\pi}  \frac{1}{u}
    P_{ff}^{(0)}(z)\; \theta_{1-z\geq \lambda e^{-t}}
\\&
 =\frac{\alpha_S^{(0)}((1-x/u)e^t)}{\pi}  \frac{1}{u}
    P_{ff}^{(0)}(x/u)\; \theta_{u-x\geq u\lambda e^{-t}}\; .
\end{split}
\end{equation}

{\em Case (C)}:  
The coupling constant $\alpha_S$ depends on 
the transverse momentum $k^T=(u-x)e^t$, while for
an IR cut-off we choose $\Delta(t,u)=\Delta(t)=\lambda e^{-t}$.
The kernel reads:
\begin{equation}
\label{eq:kernBremsC}
\begin{split}
\Kcal^{\theta(C)}_{ff} (t, x, u)
& =\frac{\alpha_S^{(0)} ((u-x)e^t)}{\pi}  \frac{1}{u}
    P_{ff}^{(0)}(x/u)\theta_{u-x\geq \lambda e^{-t}}.
\end{split}
\end{equation}

For the  (B) and (C) cases the choice of $\lambda$ for the IR cut-off must be
such that we avoid  the Landau pole in 
-- the insertion of the $1-z$ factor to 
the argument of the coupling constant
resums higher order effects in the bremsstrahlung
parts of the evolution \cite{Amati:1980ch}.
In the non-diagonal, flavour-changing,  elements of the  kernel,
 there is no need
for this kind of resummation.
We can, therefore, use $\alpha_S(e^t)$.
On the other hand, although quark gluon transition kernel elements
have neither IR divergence nor a Landau pole, it makes sense to keep
the restriction $u-x<\Delta(t,u)$,
because the cut-out part of the phase space is (will be) already
populated by the $k^T$ distribution of the primordial parton%
\footnote{Without this restriction in the early evolution time quark gluon 
    transitions would completely take over the bremsstrahlung.
}.

\subsection{Virtual part of the kernel and form-factors}
\label{sec:formfactor}

For the evolution with any type of kernel
the momentum sum rule
\begin{equation}
\begin{split}
& 0= \partial_t \sum_{f} \int dx\; x D_{f} (t, x)
   = \sum_{f} \int^1_0 d u 
   \left\{ \sum_{f'} \int^u_0 d x\;
           x\Kcal_{ff'} (t, x, u) 
  \right\} D_{f'}( t, u)
\\&
 = \sum_{f} \int^1_0 d u 
   \left\{
     -u\Kcal^v_{ff}(t,u) 
      +\sum_{f'} \int_0^u dx\; x\Kcal^\theta_{ff'}(t,x,u)
  \right\} D_{f'}( t, u)
\end{split}
\end{equation}
is imposed, the same as in the reference DGLAP case.

This sum rule determines unambiguously
the virtual part of the kernel for all cases (A--C)
\begin{equation}
  \Kcal^v_{ff}(t,u) 
= \sum_{f'} \int^u_0 
  \frac{d x}{u} x \; \Kcal^{\theta}_{f'f} (t, x, u).
\end{equation}
It should be stressed,
that in case (C) $\Kcal^v(t,u)$ includes implicitly
$\theta_{u>\Delta(t)}$, as visualized in Fig.~\ref{fig:x-u-plane}(b).
The following Sudakov form-factor results immediately:
\begin{equation}
\begin{split}
&\Phi_f (t_1, t_0 |u)
  =\int^{t_1}_{t_0} d t\; {\Kcal^v_{ff} (t, u)} 
  = \sum_{f'} \int^{t_1}_{t_0} d t 
    \int_0^u \frac{d x}{u} x\; \Kcal^\theta_{f'f}(t,x,u)
\\& 
=\sum_{f'} \int^{t_1}_{t_0} d t 
   \int_0^u \frac{d y}{u} (u-y)\; 
    \Kcal^\theta_{f'f}(t,u-y,u)
  =\sum_{f'} \int^{t_1}_{t_0} d t 
   \int_0^1 dz\; uz \Kcal^\theta_{f'f}(t,uz,u),
\end{split}
\end{equation}
where $z\equiv x/u$ and $y\equiv u-x = (1-z)u$.
The $\Kcal^\theta_{f'f}$
is constructed using the IR-singular, non-singular and
flavour-changing parts  of
the DGLAP (LL) kernel according to the following decomposition
\begin{equation}
z P_{f'f}^\theta(z)=
    \biggl[
  \delta_{f'f} \left(\frac{A_{ff}}{1-z} + F_{ff}(z) \right)
 +(1-\delta_{f'f})F_{f'f}(z)
    \biggr] \theta_{u-x>\Delta(u)}.
\end{equation}
For the list of the coefficients $A_{ff}$ and the functions $F_{f'f}(z)$
see Appendix of ref.~\cite{Golec-Biernat:2006xw}.

We also need to define the generalized kernels beyond the case of bremsstrahlung,
that is for the quark gluon transitions.
One of the possible extensions, valid for all three cases $X=A,B,C$, reads
\begin{equation}
\label{eq:kernGen}
x\Kcal^{\theta(X)}_{f'f}(t,x,u)
 = \delta_{f'f}\; x\Kcal^{\theta(X)}_{f'f}(t,x,u)
 +(1-\delta_{f'f}) \frac{\alpha_S (e^t)}{\pi} F_{f'f}(z)
     \theta_{u-x>\Delta^{(X)}(u)},
\end{equation}
where $\alpha_S$ in the flavour changing elements has no
$z$- or $k^T$-dependence and the IR cut-off $\Delta^{(X)}$ is the same as
in the bremsstrahlung case.
The Sudakov form-factors resulting from the above kernels
are split into three corresponding parts:
\begin{equation}
 \Phi_f(t_1,t_0|u) 
 =\mathbf{\Phi}_f (t_1, t_0|u)
       + \Phi^b_f (t_1, t_0|u)
       + \Phi^c_f (t_1, t_0|u).
\end{equation}
In case (C) we get three genuinely $u$-dependent components
\begin{equation}
\label{eq:PhiC}
\begin{split}
\mathbf{\Phi}_f (t_1, t_0 |u)
& =\int^{t_1}_{t_0} d t\; \int_0^1 dz\;
  \frac{\alpha_S ((1-z)ue^t)}{\pi}
  \frac{A_{ff}}{1-z}
  \theta_{(1-z)u>\lambda e^{-t}}
\\& =A_{ff} \frac{2}{\beta_0} 
      \varrho_2(\tB_0+\ln u,\tB_1 +\ln u;\tB_\lambda),
\\
\Phi^b_f (t_1, t_0 |u) 
& =\int^{t_1}_{t_0} d t\;
  \int_0^1 dz\;
  \frac{\alpha_S ((1-z)ue^t)}{\pi}
  F_{ff}(z)\;
  \theta_{(1-z)u>\lambda e^{-t}}\; ,
\\
\Phi^c_f (t_1, t_0 |u) 
&=\int^{t_1}_{t_0} d t\;
  \frac{\alpha_S (e^t)}{\pi}
  \sum_{f'\neq f} 
  \int_0^1 dz\;
  F_{f'f}(z)\;
  \theta_{(1-z)u>\lambda e^{-t}},
\end{split}
\end{equation}
where $\tB_i\equiv t_i-\ln\Lambda_0$, $\tB_\lambda\equiv t_\lambda-\ln\Lambda_0$ 
and function $\varrho_2$ is defined
in Appendix~\ref{append:trapez} in terms of log functions.
The integration domains for the consecutive
form-factors of the above type are also shown in
a pictorially way in Fig.~\ref{fig:sudak0},
using the well known logarithmic Sudakov plane.

In case (B) the $u$ dependence disappears due to the fact that 
both $\alpha_S((1-z)e^t)$ and 
$\theta_{1-z>\Delta(t)}$ depend on $u$ exclusively through $z=x/u$:
\begin{equation}
\mathbf{\Phi}_f (t_1, t_0)=\mathbf{\Phi}_f (t_1, t_0 |1),
\qquad
\Phi^b_f (t_1, t_0)=\Phi^b_f (t_1, t_0 |1),
\qquad
\Phi^c_f (t_1, t_0)=\Phi^c_f (t_1, t_0 |1).
\end{equation}
In other words, setting $u=1$ brings us from case (C) to the case (B).

The reason  behind the seemingly at hoc  three-fold split of 
$\mathbf{\Phi}_f (t_1, t_0)$ is practical.  
In the MC  the form-factor has to be calculated event-per-event.
One-dimensional integration for each MC event is acceptable --
it does not slow down MC generation noticeably.
Here, the most singular part of ${\bf \Phi}_f$ is
calculable analytically.
In $\Phi^b_f$ we are able to integrate analytically over $t$
and the integration over $z$ is done numerically
while in $\Phi^c_f$ we can integrate analytically over $z$ and the
integration over $t$ is has to be done numerically
(see also Appendix \ref{append:mapping}).
Altogether, we are thus able to avoid 2-dimensional numerical integration
for each MC event
(or use of  look-up tables and interpolation for fast 
evaluation of the Sudakov form-factors for each MC event).

Finally, the form-factors for the simplest DGLAP LL case read as follows:
\begin{equation}
\begin{split}
\mathbf{\Phi}_f (t_1, t_0)
& = \frac{2}{\beta_0}\; (\tau(t_1)-\tau(t_0))\;
      A_{ff} \ln\frac{1}{\epsilon},
\\
\Phi^b_f (t_1, t_0)
& = \frac{2}{\beta_0}\;(\tau(t_1)-\tau(t_0))
  \int_0^1 dz\; F_{ff}(z)\;
  \theta_{1-z>\epsilon}\; ,
\\
\Phi^c_f (t_1, t_0)
&=\frac{2}{\beta_0}\;(\tau(t_1)-\tau(t_0))
  \sum_{f' \neq f} 
  \int_0^1 dz\;
  F_{f'f}(z)\;
  \theta_{1-z>\epsilon}\; ,
\end{split}
\end{equation}
where $\tau(t)=\ln(t-\ln\Lambda_0)$.

At present, in the MC implementation, we use for cases (B) and (C),
a slightly different
form of the quark gluon changing kernels elements:
\begin{equation}
\label{eq:kernGen1}
\begin{split}
&x\Kcal^{\theta(B')}_{f'f}(t,x,u)
 = \delta_{f'f}\; x\Kcal^{\theta(B)}_{f'f}(t,x,u)
 +(1-\delta_{f'f}) \frac{\alpha_S ((1-z)e^t)}{\pi} F_{f'f}(z)
  \theta_{1-z>\lambda e^{-t}}\; ,
\\
&x\Kcal^{\theta(C')}_{f'f}(t,x,u)
 = \delta_{f'f}\; x\Kcal^{\theta(C)}_{f'f}(t,x,u)
 +(1-\delta_{f'f}) \frac{\alpha_S (u(1-z)e^t)}{\pi} F_{f'f}(z)
  \theta_{y>\lambda e^{-t}}\; ,
\end{split}
\end{equation}
that is, we use the same arguments of $\alpha_S$ as for gluonstrahlung.
We will refer to them as cases (B') and (C').
This is done mainly to facilitate numerical comparisons with
our Markovian MCs.
One can easily go back from cases (B') and (C') to (B) and (C)
with an extra (well behaving) MC weight, if  needed.
The corresponding form-factor $\Phi^c_f (t_1, t_0)$ gets
properly  redefined in cases (B') and (C'), of course.

\section{Constrained Monte Carlo for pure bremsstrahlung}
\label{sec:gluonstrahlung}
Let us discuss the case of pure gluonstrahlung first. We will focus
our attention
on the following integral, being part of eq.~(\ref{eq:EvolsoluG})
\begin{equation}
\label{eq:U-brems}
\begin{split}
G^n_{ff}(K^B;t_{b},t_{a},x,u) 
&=
\bigg[ \prod_{i=1}^n 
  \int_{t_{i-1}}^{t_b} dt_i\;
  \int_x^u dx_i\;
  \Keu^R_{ff} (t_i,x_i,x_{i-1})
\bigg]
e^{-\sum_{n=1}^{n+1}\Phi_{f_{i-1}}(t_i,t_{i-1}|x_{i-1})}
\delta_{x=x_n}.
  \end{split}
\end{equation}
It describes the emission of $n$ gluons.
The following ``aliasing'' of variables is used:
 $t_{n+1}\equiv t_b$, $t_{0}\equiv t_a$ and $x_0\equiv u$.
The integrand is well approximated
by the product of the IR singularities in terms of variables $ y_i=x_i-x_{i-1}$
\begin{equation}
\prod_{i=1}^n \Keu^R_{ff}(x_i,x_{i-1})
    \simeq \prod_{i=1}^n \frac{1}{x_i-x_{i-1}}
    =      \prod_{i=1}^n \frac{1}{y_i}
    \simeq \prod_{i=1}^n \frac{1}{1-z_i}.
\end{equation}
Hence, switching from  variables $x_i$ to $y_i$ or $z_i$ is almost mandatory
and the multigluon distribution with
the $\delta$-function constraining the total energy of emitted gluons
takes the following symmetric form:
\begin{equation}
\int  \delta_{x=x_n} \prod_{i=1}^n \frac{dx_i}{x_i-x_{i-1}}
=\int \delta\left(u-x-\sum_{j=1}^n y_j\right) \prod_{i=1}^n  \frac{dy_i}{y_i}
\simeq \int \delta\left(x-u\prod_{j=1}^n z_j\right) \prod_{i=1}^n \frac{dz_i}{1-z_i}.
\end{equation}
Constructing the MC program/algorithm
for the multidimensional distribution featuring such
a $\delta$-function is a hard technical problem
and it is the problem
of constructing a Constrained Monte Carlo (CMC).

It should be kept in mind,
that it is possible, as shown in ref.~\cite{Jadach:2005yq},
to generate the above distribution in $x$-space
without such a $\delta$-function, provided $G_{ff}$ is convoluted
with the power-like function $x_0^\omega$.
Such a solution was labeled as CMC class II,
while the CMC of this paper was already
referred to in ref.~\cite{Jadach:2005yq} as CMC class I.

In ref.~\cite{Jadach:2005bf} the first CMC algorithm of the class I
was found and tested for DGLAP kernel, that is for our  case (A).
This algorithm is based on the observation that for the product of steeply
rising functions, proportional to $ 1/y_i$, the $\delta$-function constraint  
is effectively resolved  by a single (let's say) $y_k$,
while all {\it other} $y_i$,
can be considered as unconstrained.
We shall extend the  CMC class I solution
to more complicated kernels, that is of our type  (B) and (C).
The main complication with respect to case (A)
is due to a more complicated singular $y$-dependence
(or $z$-dependence) entering through the coupling constant $\alpha_S$,
and even more important, through form-factors.
In addition, our new CMCs will not only generalize
the solutions of ref.~\cite{Jadach:2005bf}, but will
be described in such a way that any future extension
to other types of evolution kernels will be rather easy.
In the following sub-sections we shall present the details of 
our generalized solutions.

\subsection{Generic CMC class I}
As already indicated, we intend to introduce the formulation
of the CMC algorithm which covers three types of kernels (A--C),
and also that further extensions are possible and easy.
At first, let us consider the following generic expression including 
the sum of constrained multidimensional integrals
\begin{equation}
\label{eq:Dgener0}
D(v)= \Psi'(v)\; e^{-\int_{v_0}^{v_x} K(v') dv'}
\bigg\{\delta_{\Psi(v)=\Psi(v_0)}
  +\sum_{n=1}^\infty \frac{1}{n!}
   \bigg[ \int_{v_0}^v
   \prod_{i=1}^n K(v_i) dv_i \bigg]
   \delta_{\Psi(v)=\sum_{j=1}^n \Psi(v_j)}
\bigg\}.
\end{equation}
The following properties will be assumed:
(a)~The function $\Psi(v)$, used to change integration variables,
 must be  monotonous; 
 its derivative must remain non-negative,  $\frac{d\Psi}{dv}=\Psi'(v)\geq 0$,
for all $v>v_0$.
(b)~The positioning of the IR term%
\footnote{This term is up to a constant (choice of integration variable)
         equivalent to $\delta_{\Psi(v)=\Psi(v_0)}$.}
  $\delta(v-v_0)$ at $v=v_0$ is a convention.
  In practical MC realization it represents a no-emission event%
\footnote{The ``no-emission'' MC event will have precise interpretation,
          independently of the choice of $v_0$.}.
  Let us stress that in the $v$-space we are free to place the position $v_{IR}$
  of the IR part of spectrum $\delta(v-v_{IR})$ anywhere outside
  the $(v_0,v_x)$ interval; we have opted for $v_{IR}=v_0$
(c)~the variable $v_x$ controlling the overall normalization will be specified only 
  later; it will be adjusted to get convenient normalization,
(d)~the true upper integration limit of $v_i$ is below $v$
  and is in fact uniquely determined by the $\delta$-function of the constraint.

In the following examples,  the variable $v_i$ will be defined as
$v_i=\ln y_i=\ln x_i-x_{i-1}$
or $v_i=\ln(1-z_i)= \ln(1-x_i/x_{i-1})$, while the function $\Psi(v)$
will be typically rather simple; $\Psi(y_i)=y_i$ or $\Psi(z_i)=\ln z_i$.

Assuming $K(v)> 0$, we can define a mapping (and its inverse)
which removes $K(v)$ from the integrand:
\begin{equation}
 r_i=R(v_i)=\int_{v_0}^{v_i} K(v') dv',\qquad
 v_i=V(r_i)=R^{-1}(r_i).
\label{eq:cum-var}
\end{equation}
Our master formula transforms then as follows:
\begin{equation}
D(v)= \Psi'(v) e^{-R(v_x)}
\bigg\{\delta_{\Psi(v)=\Psi(v_0)}
  +\sum_{n=1}^\infty \frac{1}{n!}
   \bigg[ \int_{0}^{R(v)}
   \prod_{i=1}^n dr_i \bigg]
   \delta_{\Psi(v)=\sum_{j=1}^n \Psi(R^{-1}(r_j))}
\bigg\}.
\end{equation}
The function $f(r)=\Psi(R^{-1}(r))$ is usually a very steeply growing 
function of $r$, hence the constraint is effectively resolved 
by a single $r_j\simeq R(v)$, the biggest one.
Let us exploit this fact in order to replace
the complicated constraint with the simpler one%
\footnote{The function $\max_j r_j$ is equal the biggest  $r_j$ among
   $j=1,2,...,n$.}
$\delta(R(v)-\max_j r_j)$.
This is done in three steps.
{\em Step one:} introduce a new 
auxiliary integration variable $X$ countered by a $\delta$-function:
\begin{equation}
\begin{split}
D(v)=& 
e^{-R(v_x)}\delta_{v=v_0}
  +\Psi'(v) e^{-R(v_x)}
\\&\times
   \sum_{n=1}^\infty \frac{1}{n!}
   \int d X
   \bigg[ \int_{0}^{R(v)}
   \prod_{i=1}^n dr_i \bigg]
   \delta_{\Psi(v)=\sum_{j=1}^n \Psi(R^{-1}(r_j))}
   \delta_{R(v)=X+\max_j r_j}.
\end{split}
\end{equation}
{\em Step two:} change the variables $r_i=r'_i-X$:
\begin{equation}
\begin{split}
D(v)=& 
e^{-R(v_x)}\delta_{v=v_0}
  +\Psi'(v) e^{-R(v_x)}
\\&\times
   \sum_{n=1}^\infty \frac{1}{n!}
   \int d X
   \bigg[ \int_{0}^{R(v)}
   \prod_{i=1}^n dr'_i \; \theta_{r_i>0} \bigg]
   \delta_{\Psi(v)=\sum_{j=1}^n \Psi(R^{-1}(r'_j-X))}
   \delta_{R(v)=\max_j r'_j}.
\end{split}
\end{equation}
From now on we have to watch out for the condition $r_i=r'_i-X>0$ explicitly%
\footnote{On the other hand, no problem with $r'_i\leq v$, 
   because we shall get $X\geq 0$.}.
{\em Step three:} eliminate the old constraint by integrating over $X$
\begin{equation}
D(v)=
e^{-R(v_x)}\delta_{v=v_0}
  +e^{-R(v_x)}
   \sum_{n=1}^\infty \frac{1}{n!}
   \bigg[ \int_{0}^{R(v)}
   \prod_{i=1}^n dr'_i \; \theta_{r_i>0} \bigg]
   \delta_{R(v)=\max_j r'_j}\; \mathfrak{J},
\end{equation}
where $X_0(r'_1,r'_2,...,r'_n)$ is found by means of
solving numerically (iterative method)
the original constraint for $X$.
The Jacobian factor $\mathfrak{J}$ is defined as follows
\begin{equation}
 \mathfrak{J}=
 \frac{\Psi'(v)}{\partial_X \sum_{j=1}^n \Psi(R^{-1}(r'_j-X_0)) }
=\frac{\Psi'(v)}{\sum_{j=1}^n \Psi'(v_j) (R'(v_j))^{-1} }.
\end{equation}
In the above factor the $j$-th term satisfying $R(v)=\max_j r'_j$
dominates the sum in the denominator; consequently $\mathfrak{J}\simeq R'(v)=K(v)$.
This  form, valid in the limit only,  we keep explicitly
in the integrand, while the remaining part of $\mathfrak{J}$
is incorporated in a complicated but mild Monte Carlo weight:
\begin{equation}
D(v)=
e^{-R(v_x)}\delta_{v=v_0}
  +R'(v) e^{-R(v_x)}
   \sum_{n=1}^\infty \frac{1}{n!}
   \bigg[ \int_{0}^{R(v)}
   \prod_{i=1}^n dr'_i \bigg]
   \delta_{\max_j r'_j=R(v)}\; w^\#,
\end{equation}
where the MC weight is
\begin{equation}
\label{eq:w-hash}
  w^\# =\frac{\Psi'(v)}{R'(v)\sum_{j=1}^n \Psi'(v_j) (R'(v_j))^{-1} }
      \prod_{i=1}^n \theta_{r_i>0}.
\end{equation}

Last but not least, we isolate cleanly a part of 
the integrand normalized properly to 1
with the help of the integration variable $\xi_i=r_i/R(v)$
and the Poisson distribution $P(n|\lambda)=\exp(-\lambda) \lambda^n/n!$:
\begin{equation}
D(v)=
  e^{-R(v_x)}\delta_{v=v_0}
  +R'(v) e^{R(v)-R(v_x)}
   \sum_{n=1}^\infty
   P(n-1|R(v))
   \bigg[
   \prod_{i=1}^n  \int_{0}^{1}d\xi_i \bigg]
   \frac{\delta_{\max_j \xi_j}=1}{n}\;
   w^\#(\xi),
\end{equation}
The nice thing is that $\bar{D}(v)$, obtained by neglecting $w^\#$,
is known analytically
\begin{equation}
\bar{D}(v)=D(v)|_{w^\#=1}
   =e^{-R(v_x)}\delta_{v=v_0}
   +\theta_{v>v_0}R'(v) e^{R(v)-R(v_x)}
\end{equation}
and is normalized to 1:
\begin{equation}
\int_{v_0}^{v_x} \bar{D}(v) dv
= e^{-R(v_x)} + \int_{\exp(-R(v_x))}^1 d\Big( e^{R(v)-R(v_x)} \Big)
= e^{-R(v_x)} +\Big(1-e^{-R(v_x)}\Big)
=1.
\end{equation}
The above convenient unitary normalization is achieved by means of identifying
$v_x$ with the upper limit of $v$, see also below for particular realizations.
In the implementation of quark gluon transitions using  FOAM~\cite{foam:2002}
(see section~\ref{sec:completeQG})
one introduces the integration variable $U=U(v)=e^{R(v)-R(v_x)}\in (0,1)$
and the above equation transforms into
\begin{equation}
\label{eq;Uintegral}
 \int_0^{\exp(-R(v_x))} dU + \int_{\exp(-R(v_x))}^1 dU
 =\int_0^1 dU.
\end{equation}

The MC procedure of generating the variables $v$, $n$ and $(v_1,v_2,...,v_n)$
obeying the constraint is the following:
\begin{itemize}
\item
Generate $v$ according to $\bar{D}(v)$ times whatever
the other function of $v$ in the Monte Carlo problem,
for that purpose use the variable $U=e^{R(v)-R(v_x)}\in(0,1)$.
\item
If $U\leq e^{-R(v_x)}$ then set $v=v_0$
and $n=0$ (no-emission event).
\item
Otherwise $U$ is translated into $v$ ($v>v_0$)
and $n>0$ is generated according to $P(n-1|R(v))$.
\item
Generate the variables $\xi_i\in (0,1), i=1,2,...,n$,
except one of them: $\xi_j=1$,
where $j=1,2,...n$ is chosen randomly with the uniform probability.
\item
The variables $\xi_i$ are now translated into $r'_i$ and
the transcendental equation defining the shift $X_0$ is solved numerically.
\item
Once $X_0$ is known, then all $v_i(r_i(r'_i((\xi_i)))$ are calculated.
\item
The MC weight $w^\#$ is evaluated;
the check on $r_i>0$ can be done earlier.
\end{itemize}
The above algorithm generates weighted MC events $(n;v_1,v_2,...,v_n)$
exactly according to the resumed series of the integrands defining $D(v)$.

\subsection{Treatment of $t$-ordering -- generic case}
\label{sec:torder}
In case (A) of the DGLAP kernel,
discussed in ref.~\cite{Jadach:2005bf},
the $t$-ordering in the integrals of eq.~(\ref{eq:U-brems})
can be easily traded for a $1/n!$ factor, while
$K(v)$ in the previous section includes un-ordered
integrals $\int dt_i$ over the entire available range for
each $t_i$.
However, even in this simple case one has to be careful if one aims not
only at the numerical evaluation of the PDF, but also at the proper simulation
of the entire MC events
$\{n,(t_1,x_1),(t_2,x_2),...,(t_n,x_n)\}$.
The CMC class I algorithm of ref.~\cite{Jadach:2005bf}
(DGLAP) is formulated in terms of $z_i=x_i/x_{i-1}$;
the point is that at the end of the MC algorithm
one has to calculate $x_i$ using a well defined
ordering of $z_i$, which is exactly the same as in the sequence
of the (originally) un-ordered $t_i$.
In practice, in the end of the MC generation,
one has to order $t_i$ and $z_i$
{\em simultaneously} before calculating $x_i=x_0 z_1 z_2...z_i$.
This is because the original integrand is symmetric with
respect to interchange of the pairs
of variables $(t_i,z_i)\leftrightarrow (t_k,z_k)$
and not with respect to interchanging $t_i\leftrightarrow t_k$
or $z_i\leftrightarrow z_k$ done independently.
This property we will call the {\em pairwise} permutation symmetry.

In our most general case (C), the integrand is not {\em pairwise} symmetric,
mainly due to nontrivial $x_i$-dependency
in the form-factor, see eq.~(\ref{eq:U-brems}).
In this case, the strategy is such that we introduce a simplified
integrand, with the simplified form-factor and simplified kernel
$\bar{\Pbbm}$
(the IR-singular part of the kernel), such that the simplified
integrand does feature the {\em pairwise} permutation symmetry.
The above simplifications are
immediately and exactly compensated by means of the
MC weight $w^{\mathbbm{P}}$,
which absorbs all possible {\em pairwise} non-symmetry.

Let us translate what has been said above into rigorous algebra.
We start from a more sophisticated variant of our
generic multi-integral (\ref{eq:Dgener0}),
covering all cases (A--C) and possibly other similar cases,
but this time including explicitly ordered $t$-integrations:
\begin{equation}
\label{eq:generic2}
\begin{split}
D(t,t_0|v)=& \Psi'(v)
e^{-\int_{t_0}^t dt'
    \int_{v_0(t')}^{v_x} \bar{\Pbbm}(v',t') dv'}
\bigg\{\delta_{\Psi(v)=\Psi(v_0)}+
\\&
  +\sum_{n=1}^\infty
   \bigg[
   \prod_{i=1}^n
   \int_{t_{i-1}}^t dt_i\;
   \int_{v_0(t_i)}^v dv_i\;
   \bar{\Pbbm}(t_i,v_i)\bigg]
   w^{\mathbbm{P}}({\bf t},{\bf v})\;
   \delta_{\Psi(v)=\sum_{j=1}^n \Psi(v_j)}
\bigg\},
\end{split}
\end{equation}
where $t_n\equiv t$. The bold-face variable ${\bf t}$ denotes 
the entire vector $(t_1,t_2,...,t_n)$ and
similar convention  is used  for the definition of the vector ${\bf v}$.

In order to get rid of the $t$-ordering in the basic MC
algorithm we proceed carefully step by step as follows:
\begin{enumerate}
\item
We introduce formally {\em pairwise} symmetrization, i.e. we sum
up over all permutations $\Pmf$ of the pairs of the variables
$(v_i,t_i)$, compensating by means of the $1/n!$ factor:
\begin{displaymath}
\frac{1}{n!} \sum_{\Pmf}
\bigg[
  \prod_{i=1}^n
  \int_{t_0}^t dt^{\Pmf}_i\;
  \int_{v_0(t^{\Pmf}_i)}^v dv^{\Pmf}_i\;
  \bar{\Pbbm}(t^{\Pmf}_i,v^{\Pmf}_i)
\bigg]
  \theta_{t^{\Pmf}_n>t^{\Pmf}_{n-1}>... >t^{\Pmf}_{1}}\;
  w^{\mathbbm{P}}({\bf t}^{\Pmf},{\bf v}^{\Pmf}).
\end{displaymath}
\item
The part of the integrand in the brackets is now perfectly
{\em pairwise} symmetric and the permutation $\Pmf$ can be undone in this part:
\begin{displaymath}
\frac{1}{n!}
\bigg[
  \prod_{i=1}^n
  \int_{t_0}^t dt_i\;
  \int_{v_0(t_i)}^v dv_i\;
  \bar{\Pbbm}(t_i,v_i)
\bigg]
\sum_{\Pmf}
  \theta_{t^{\Pmf}_n>t^{\Pmf}_{n-1}>... >t^{\Pmf}_{1}}\;
  w^{\mathbbm{P}}({\bf t}^{\Pmf},{\bf v}^{\Pmf}).
\end{displaymath}
\item
The sum over permutations
$\sum_{\Pmf}$ can be dropped out because only one
permutation contributes at a given point ${\bf t}=(t_i,t_2,...,t_n)$.
This particular permutation we denote by $\Pmf_{\bf t}$, obtaining:
\begin{equation}
\label{eq:Dgener1}
\begin{split}
D(t,t_0|&v)= \Psi'(v)
e^{-\int_{t_0}^t dt'\int_{v_0(t')}^{v_x} \bar{\Pbbm}(v',t') dv'}
\bigg\{\delta_{\Psi(v)=\Psi(v_0)}+
\\&
+\sum_{n=1}^\infty  
 \bigg[
   \prod_{i=1}^n
   \int_{t_{i-1}}^t dt_i\;
   \int_{v_0(t_i)}^v dv_i\;
   \bar{\Pbbm}(t_i,v_i)
  \bigg]
   \delta_{\Psi(v)=\sum_{j=1}^n \Psi(v_j)}\;
   w^{\mathbbm{P}}({\bf t}^{\Pmf_{\bf t}},{\bf v}^{\Pmf_{\bf t}})
\bigg\}.
\end{split}
\end{equation}
\item
In the basic MC the weight $w^{\mathbbm{P}}$ is temporarily neglected and
$t_i$ are generated unordered. The permutation $\Pmf_{\bf t}$
is then read from the ordering in ${\bf t}=(t_1,t_2,...,t_n)$.
It is then used to construct
the sequence of $x_i$ out of $v_i$ and to calculate $w^{\mathbbm{P}}$.
\end{enumerate}

In order to finally bring eq.~(\ref{eq:Dgener1}) into the
standardized form of eq.~(\ref{eq:Dgener0}) we interchange
the order of integration over $t_i$ and $v_i$
\begin{displaymath}
   \int_{t_{i-1}}^t dt_i\;
   \int_{v_0(t_i)}^v dv_i\;\bar{\Pbbm}(t_i,v_i)
   =
   \int_{v_0}^v dv_i\;
   \int_{t_{i-1}(v_i)}^{t(v_i)} dt_i\;\bar{\Pbbm}(t_i,v_i)
  = \int_{v_0}^v dv_i\; K(v_i) \int_0^1 d\sigma_i.
\end{displaymath}
The additional change of variables
\begin{equation}
\label{eq:sigma}
   \sigma(t,v)
 =\frac{\int_{t_{\min}(v)}^{t} dt'\;\bar{\Pbbm}(t',v)}%
       {\int_{t_{\min}(v)}^{t_{\max}(v)} dt'\;\bar{\Pbbm}(t',v)}
 =\frac{\int_{t_{\min}(v)}^{t} dt'\;\bar{\Pbbm}(t',v)}%
       {K(v)}
\end{equation}
allows to map, for every value of $v_i$,
 uniformly distributed $\sigma_i$ into $t_i$.
For a particular realization see Appendix~\ref{append:mapping}.
Luckily, $\sigma_i(t_i)$ can be inverted analytically,
i.e. $t_i(\sigma_i,v_i)$ is available as an analytical formula,
for all cases (A--C).

The main purpose of the sub-section, was to obtain
the following, ready for the MC implementation, representation
of our standardized generic formula:
\begin{equation}
\label{eq:Dgener2}
\begin{split}
&D(t,t_0|v)= \Psi'(v)
e^{-\int_{v_0}^{v_x} K(v') dv'}
\bigg\{\delta_{\Psi(v)=\Psi(v_0)}+
\\&~~~~~~~~~~~~~
+\sum_{n=1}^\infty\frac{1}{n!}
 \bigg[
   \prod_{i=1}^n
   \int_{t_{i-1}}^t dt_i\;
   K(v_i)
   \int_0^1 d\sigma_i\;
  \bigg]
   \delta_{\Psi(v)=\sum_{j=1}^n \Psi(v_j)}\;
   w^{\mathbbm{P}}({\bf t}^{\Pmf_{\bf t}},{\bf v}^{\Pmf_{\bf t}})
\bigg\}
\\&~~
= e^{-R(v_x)}\delta_{v=v_0}
  +\theta_{v>v_0}
   R'(v) e^{R(v)-R(v_x)}
\\&~~~~~~~~~~~~\times
   \sum_{n=1}^\infty
   P(n-1|R(v))
   \bigg[
      \prod_{i=1}^n  \int_{0}^{1}d\xi_i \int_0^1 d\sigma_i\;
   \bigg]
   \frac{\delta_{\max_j \xi_j}=1}{n}\;
   w^\#(\mathbf{\xi})\;
   w^{\mathbbm{P}}({\bf t}^{\Pmf_{\bf t}},{\bf v}^{\Pmf_{\bf t}}).
\end{split}
\end{equation}
In the following subsection we shall describe three
realizations of the above CMC class I schemes for
three types of the kernels, (A--C).

\subsection{CMC case (A), DGLAP}
Let us start with the particular realization of the above CMC
for the easiest case of the DGLAP kernel, type (A).
Such a CMC was exposed in detail already in ref.~\cite{Jadach:2005bf}.
Here it serves as a reference case and a warm-up example.
We recall the pure bremsstrahlung evolution operator
of eq.~(\ref{eq:EvolsoluG}) in a  form adopted for further manipulations:
\begin{equation}
\begin{split}
\frac{x}{u}
G_{ff}(K^B;t_{b},t_{a},x,u) 
  &= e^{-\Phi_f(t_b,t_a)} \delta_{x=u}
  +\sum_{n=1}^\infty \;
      \bigg[ \prod_{i=1}^n 
      \int_{t_a}^{t_b} dt_i\; \theta_{t_i>t_{i-1}}  
      \int_x^u dx_i
      \bigg]
\\&\times
   e^{-\Phi_f(t_b,t_n)}
   \bigg[\prod_{i=1}^n 
        \frac{x_i}{x_{i-1}}
        \Keu^R_{ff} (t_i,x_i,x_{i-1}) 
         e^{-\Phi_{f}(t_i,t_{i-1})} \bigg]
   \delta_{x=x_n},
  \end{split}
\end{equation}
where we have dropped the non-existing dependence on $x_i$ in the form-factor $\Phi$.
Hence, one may exploit the relation
$\sum_{i=0}^n\Phi(t_i,t_{i-1})=\Phi(t_b,t_a)$.
After identification of terms and change of integration variables
\[
  x_i\Keu^R_{ff} (t_i,x_i,x_{i-1})
 =\frac{\alpha_S(e^{t_i})}{\pi} z_i P_{ff}(z_i)\theta_{1-z_i>\epsilon},
  \qquad z_i=x_i/x_{i-1},
\]
a simplified expression is obtained:
\begin{equation}
\begin{split}
&\frac{x}{u}
  G_{ff}(K^B;t_{b},t_{a},x,u)
  = e^{-\Phi_f(t_b,t_a)} \delta_{x=u}+
\\&~~~~~~
  +e^{-\Phi_f(t_b,t_a)}
   \sum_{n=1}^\infty \;
      \bigg[ \prod_{i=1}^n 
      \int^{t_b}_{t_{i-1}} dt_i\;
      \int^1_{x/u} dz_i\;
      \frac{\alpha_S(e^{t_i})}{\pi} z_iP_{ff} (z_i)
      \theta_{1-z_i>\epsilon}
   \bigg]
   \delta_{x=u\prod_{j=1}^n z_j}.
  \end{split}
\end{equation}
In the next step, the simplified kernel is introduced
\begin{equation}
\begin{split}
&\prod_{i=1}^n 
    \int\limits^{1-\epsilon} dz_i\;
    \alpha_S(e^{t_i})
    z_iP_{ff} (z_i)
 \rightarrow \prod_{i=1}^n 
    \int\limits^{1-\epsilon} \frac{dz_i}{1-z_i}\;
    \alpha_S(e^{t_i}) A_{ff}=
\\&~~~~~~~~~~~~~~
=\prod_{i=1}^n 
    \int\limits_{\ln\epsilon} dv_i\;
     \alpha_S(e^{t_i}) A_{ff}
= \prod_{i=1}^n 
  \int\limits_{v_0} dv_i\; \bar{\Pbbm}(t_i,v_i),
\\&
\bar{\Pbbm}(t_i,v_i) 
  = \frac{\alpha_S(e^{t_i})}{\pi} A_{ff},
\quad v_i=\ln(1-z_i),
\quad v_0=\ln\epsilon.
  \end{split}
\end{equation}
Once  $v$ is  defined,
we are ready to deduce our kernel $K(v)$ and
constraint function $\Psi(v)$ from
\begin{equation}
\begin{split}
&\frac{x}{u}
  G_{ff}(K^B;t_{b},t_{a},x,u)
  = e^{-\Phi_f(t_b,t_a)} \frac{1}{x}\delta_{\ln(x/u)=0}
\\&~~~~~~
  +e^{-\Phi_f(t_b,t_a)}
  \sum_{n=1}^\infty \;
      \bigg[ \prod_{i=1}^n 
      \int\limits^{t_b}_{t_{i-1}} dt_i\;
      \int\limits_{v_0}^{\ln(1-x/u)} dv_i\; 
      \bar{\Pbbm}(t_i,v_i)
      \bigg]
   \frac{1}{x}\delta_{\ln (x/u) =\sum_{j=1}^n \ln(1-\exp(v_j))}\;
   w_1^{\bar{\Pbbm}},
\end{split}
\end{equation}
where
\begin{equation}
  w_1^{\bar{\Pbbm}}=\prod_{i=1}^n 
   \frac{\alpha_S(e^{t_i}) z_i(1-z_i)P_{ff} (z_i)}{\bar{\Pbbm}(t_i,v_i)}.
\label{eq:wp-A}
\end{equation}
By means of comparison of the above expressions with eq.~(\ref{eq:generic2}),
see also eqs. (\ref{eq:cum-var},\ref{eq:Dgener2}),
one can  identify the components:
\begin{equation}
\begin{split}
&v=\ln(1-x/u),\quad \Psi(v)=\ln(1-e^v),
\quad |\Psi'(v)|=\frac{e^v}{1-e^v}=\frac{u-x}{x},
\\
&K(v)=\int_{t_a}^{t_b}dt\; \bar{\Pbbm}(t_i,v_i)
     = A_{ff}\frac{2}{\beta_0}\big(\tau(t_b)-\tau(t_a)\big).
\end{split}
\label{eq:Kv-A}
\end{equation}
The expression for the basic form-factor can be identified also:
\begin{equation}
 R(v)= \int_{v_0}^{v}dv'\; K(v')
     =\int_{v_0}^{v}dv'\; \int_{t_a}^{t_b}dt'\; \bar{\Pbbm}(t',v')
     = A_{ff}\frac{2}{\beta_0}\big(\tau(t_b)-\tau(t_a)\big)(v-v_0).
\label{eq:Rv-A}
\end{equation}
The relation $v=\ln(1-x/u)$ is valid for $v>v_0$ only.
The variable $v_x$ represents the upper boundary of $v=\ln(1-x/u)$,
hence or fixed $x$ and maximal  $u=1$ we obtain $v_x=\ln(1-x)$.
With all the above elements at hand,
we are able to complete the following standardized,
accordingly to conventions of eq.~(\ref{eq:Dgener2}), formula
\begin{equation}
\begin{split}
&\frac{x}{u}
  G_{ff}(K^B;t_{b},t_{a},x,u)
= 
   \bigg\{
   e^{-R(v_x)}\delta_{v=v_0}   
  +\theta_{v>v_0}
   \frac{1}{x \Psi'(v)} R'(v) e^{R(v)-R(v_x)}
\\&~~\times
   \sum_{n=1}^\infty
   P(n-1|R(v))
   \bigg[
      \prod_{i=1}^n  \int_{0}^{1}d\xi_i \int_0^1 d\sigma_i\;
   \bigg]
   \frac{\delta_{\max_j \xi_j=1}}{n}\;
   w^\#(\mathbf{\xi})\;
   w^{\mathbbm{P}}({\bf t}^{\Pmf_{\bf t}},{\bf v}^{\Pmf_{\bf t}})
   \bigg\}.
\end{split}
\end{equation}
where
\begin{equation}
  w^{\bar{\Pbbm}}= e^{R(v_x)-\Phi_f(t_b,t_a)}
   \prod_{i=1}^n 
   \frac{\alpha_S(e^{t_i}) z_i(1-z_i)P_{ff} (z_i)}{\bar{\Pbbm}(t_i,v_i)}
\end{equation}
and the second weight $w^\#$
is fully determined from eq.~(\ref{eq:w-hash}),
supplemented with $\Psi(v)$ and $K(v)$ of eq.~(\ref{eq:Kv-A}).

In the CMC, we usually integrate over $u$ for fixed $x$,
hence the following formula is relevant
\begin{equation}
\label{eq:GintegA}
\begin{split}
&\int_x^1 du
  G_{ff}(K^B;t_{b},t_{a},x,u)
= \int_0^1 dU\; \bigg(\frac{u}{x}\bigg)^2
   \bigg\{
   \theta_{U\leq\exp(-R(v_x))}|_{v=v_0}
  +\theta_{U>\exp(-R(v_x))} 
\\&~~\times
   \sum_{n=1}^\infty
   P(n-1|R(v))
   \bigg[
      \prod_{i=1}^n  \int_{0}^{1}d\xi_i \int_0^1 d\sigma_i\;
   \bigg]
   \frac{\delta_{\max_j \xi_j=1}}{n}\;
   w^\#(\mathbf{\xi})\;
   w^{\mathbbm{P}}({\bf t}^{\Pmf_{\bf t}},{\bf v}^{\Pmf_{\bf t}})
   \bigg\}.
\end{split}
\end{equation}
where%
\footnote{Another ingredient was the identity
  $\frac{R'(v)}{\Psi'(v)}=\big(\frac{d\Psi}{du}\big)^{-1} \frac{dR}{du}
  = \frac{1}{x}\frac{dR}{du}$.}
$U=U(x,u)=e^{R(v)-R(v_x)}=e^{R(\ln(1-x/u))-R(\ln(1-x))}$.
Due to relation $R(v_0)=0$, the point $v=v_0$ (IR boundary)
translates into $U_0=e^{-R(v_x)}=e^{-R(\ln(1-x))}$.
With all elements needed in eq.~(\ref{eq:Dgener2}) at hand,
we are ready to reconstruct from our generic formulation
the complete  CMC class I algorithm  of ref.~\cite{Jadach:2005bf},
at least for pure bremsstrahlung.

\subsection{CMC case (B), $\alpha_S(e^t(1-z))$}
This case is to some extent similar to the previous DGLAP
case. We shall, therefore, concentrate on the differences.
The starting point is now
\begin{equation}
\begin{split}
\frac{x}{u}
G_{ff}(K^B;t_{b},t_{a},x,u) 
  &= e^{-\Phi_f(t_b,t_a|1)} \delta_{x=u}
  +\sum_{n=1}^\infty \;
      \bigg[ \prod_{i=1}^n 
      \int_{t_a}^{t_b} dt_i\; \theta_{t_i>t_{i-1}}  
      \int_x^u dx_i
      \bigg]
\\&\times
   e^{-\Phi_f(t_b,t_n|1)}
   \bigg[\prod_{i=1}^n 
        \frac{x_i}{x_{i-1}}
        \Keu^R_{ff} (t_i,x_i,x_{i-1}) 
         e^{-\Phi_{f}(t_i,t_{i-1}|1)} \bigg]
   \delta_{x=x_n},
  \end{split}
\end{equation}
where one may again combine form-factors into a single one: 
$\sum_{i=0}^n\Phi(t_i,t_{i-1}|1)=\Phi(t_b,t_a|1)$.
Now, in terms of
\[
  x_i\Keu^R_{ff} (t_i,x_i,x_{i-1})=\frac{\alpha_S(e^{t_i}(1-z_i))}{\pi} z_i P_{ff}(z_i)
\theta_{1-z_i>\lambda e^{-t_i}},
  \qquad z_i=x_i/x_{i-1},
\]
the simplified expression reads
\begin{equation}
\begin{split}
&\frac{x}{u}
  G_{ff}(K^B;t_{b},t_{a},x,u)
  = e^{-\Phi_f(t_b,t_a|1)} 
\bigg\{
\delta_{x=u}+
\\&~~
   +\sum_{n=1}^\infty \;
      \bigg[ \prod_{i=1}^n 
      \int^{t_b}_{t_{i-1}} dt_i\;
      \int^1_{x/u} dz_i\;
      \frac{\alpha_S((1-z_i)e^{t_i})}{\pi} z_iP_{ff} (z_i)
      \theta_{1-z_i>\lambda e^{-t_i}}
   \bigg]
   \delta_{x=u\prod_{j=1}^n z_j}
  \bigg\}.
  \end{split}
\end{equation}
Simplification of the kernel goes as follows:
\begin{equation}
\begin{split}
&\prod_{i=1}^n 
    \int\limits^{1-\lambda e^{-t_{i}}} dz_i\;
    \alpha_S((1-z_i)e^{t_i})
    z_iP_{ff} (z_i)
 \rightarrow \prod_{i=1}^n 
    \int\limits^{1-\lambda e^{-t_{i}}} 
    \frac{dz_i}{1-z_i}\;
    \alpha_S((1-z_i)e^{t_i}) A_{ff}=
\\&~~~~~~~~~~~~~~
=\prod_{i=1}^n 
    \int\limits_{\ln\lambda-t_{i}} dv_i\;
     \alpha_S(e^{t_i+v_i}) A_{ff}
= \prod_{i=1}^n 
  \int\limits_{v_0(t_i)} dv_i\; \bar{\Pbbm}(t_i,v_i),
\end{split}
\end{equation}
where
\begin{equation}
\bar{\Pbbm}(t_i,v_i) 
  = \frac{\alpha_S(e^{t_i+v_i})}{\pi}A_{ff},
\quad v_i=\ln(1-z_i)\quad \hbox{\rm and}
\quad v_0(t)=\ln\lambda - t.
\end{equation}
So far we have followed closely the DGLAP case,
except of the more complicated IR cut-off and the extra factor in the argument
of $\alpha_S$.
The choice of the variable $v$ is the same, hence also the function $\Psi(v)$
remains unchanged:
\begin{equation}
\begin{split}
&\frac{x}{u}
  G_{ff}(K^B;t_{b},t_{a},x,u)
  = e^{-\Phi_f(t_b,t_a)} \frac{1}{x}
  \bigg\{
  \delta_{\ln(1-e^v)=0}+
\\&~~~~~~~~~~~~~~~~~~
  +\sum_{n=1}^\infty \;
      \bigg[ \prod_{i=1}^n 
      \int\limits^{t_b}_{t_{i-1}} dt_i\;
      \int\limits_{v_0(t_i)}^{v} dv_i\; 
      \bar{\Pbbm}(t_i,v_i)
      \bigg]
   \delta_{\ln(1-e^v) =\sum_{j=1}^n \ln(1-e^{v_j})}\;
   w_1^{\bar{\Pbbm}}
   \bigg\},
\end{split}
\end{equation}
where
\begin{equation}
  w_1^{\bar{\Pbbm}}=\prod_{i=1}^n 
   \frac{\alpha_S(e^{t_i+v_i}) z_i(1-z_i)P_{ff} (z_i)}{\bar{\Pbbm}(t_i,v_i)}.
\label{eq:wp-B}
\end{equation}
While comparing the above expressions with eq.~(\ref{eq:generic2})
we  immediately identify the following components:
\begin{equation}
v=\ln(1-x/u),\quad \Psi(v)=\ln(1-e^v),
\quad |\Psi'(v)|=\frac{e^v}{1-e^v}=\frac{u-x}{x}.
\end{equation}
In principle, the evaluation of $K(v)$ requires a
change of the integration order used in the form-factor integral,
see also Appendix~\ref{append:mapping},
\begin{equation}
 R(v;t_b,t_a)
  =\int\limits_{t_a}^{t_b}dt'\; 
   \int\limits_{\ln\lambda-t'}^{v}dv'\;\bar{\Pbbm}(t',v')
  =\int\limits_{v_0(t_b)}^{v}dv'\!\!\!
   \int\limits_{\max(t_a,\ln\lambda -v')}^{t_b}\!\!\! dt'\; 
   \bar{\Pbbm}(t',v')
  =\int\limits_{v_0(t_b)}^{v}dv'\; K(v').
\end{equation}
In practice, it is slightly easier to calculate $R(v;t_b,t_a)$
with the $t$-integration external, and then obtain $K(v)$ by
differentiation:
\begin{equation}
 R(v;t_b,t_a)=
   A_{ff}\frac{2}{\beta_0}\; 
   \int\limits_{t_a}^{t_b}dt'\; 
   \int\limits_{\ln\lambda-t'}^{v}dv'
   \frac{1}{t'+v'-\ln\Lambda_0}
  =A_{ff}\frac{2}{\beta_0} \varrho_2(\tB_b+v,\tB_a+v;\tB_\lambda),
\label{eq:Rv-B}
\end{equation}
and
\begin{equation}
  K(v) = \partial_v R(v;t_b,t_a)
  =A_{ff}\frac{2}{\beta_0} \partial_v \varrho_2(\tB_b+v,\tB_a+v;\tB_\lambda).
\label{eq:Kv-B}
\end{equation}
The functions $\varrho_2$ and $\partial_v\varrho_2$ are given 
in appendices~\ref{append:trapez} and \ref{append:mapping}.
The rest of the algebra is almost the same as in the case of DGLAP:
\begin{equation}
\begin{split}
&\frac{x}{u}
  G_{ff}(K^B;t_{b},t_{a},x,u)
=
   \bigg\{
   e^{-R(v_x)}\delta_{v=v_0}
  +\theta_{v>v_0}  \frac{1}{x} \frac{1}{\Psi'(v)}
   R'(v) e^{R(v)-R(v_x)}
\\&~~\times
   \sum_{n=1}^\infty
   P(n-1|R(v))
   \bigg[
      \prod_{i=1}^n  \int_{0}^{1}d\xi_i \int_0^1 d\sigma_i\;
   \bigg]
   \frac{\delta_{\max_j \xi_j=1}}{n}\;
   w^\#(\mathbf{\xi})\;
   w^{\mathbbm{P}}({\bf t}^{\Pmf_{\bf t}},{\bf v}^{\Pmf_{\bf t}})
   \bigg\}.
\end{split}
\end{equation}
The variable $v_x=\ln(1-x)$ and the weights $w^\#$ and $w^{\mathbbm{P}}$ are defined
in the same way as discussed already in the case (A) of DGLAP. One has to
remember only that $v_i$ enters into $\alpha_S(e^{t_i+v_i})$.
The final integral form coincides with
that for DGLAP, see eq.~(\ref{eq:GintegA}).
The important differences are in the definitions of the components,
in particular, the function $R(v)$ is more complicated here.

\subsection{CMC case (C), $\alpha_S(k^T)$, CCFM}
For this case we are dealing with the most general implementation of the
generic solution defined in eq.~(\ref{eq:EvolsoluG}):
\begin{equation}
\begin{split}
\frac{x}{u}
&G_{ff}(K^B;t_{b},t_{a},x,u) 
  = e^{-\Phi_f(t_b,t_a|x)} \delta_{x=u}
\\&~~~
  +\sum_{n=1}^\infty \;
      \bigg[ \prod_{i=1}^n 
      \int_{t_a}^{t_b} dt_i\; \theta_{t_i>t_{i-1}}  
      \int_x^u dx_i
      \frac{x_i}{x_{i-1}}
      \Keu^R_{ff} (t_i,x_i,x_{i-1}) 
   \bigg]
   e^{-\sum_{j=0}^n\Phi_f(t_j,t_{j-1}|x_{j-1})}
   \delta_{x=x_n},
  \end{split}
\end{equation}
where we are aliasing the following variables:
$x=x_n$, $u=x_0$, $t_0=t_a$ and $t_n=t_b$.
The kernel is again more complicated than the one used in the two
previous sub-sections
\[
  x_i\Keu^R_{ff} (t_i,x_i,x_{i-1})
   =\frac{\alpha_S(e^{t_i}(1-z_i)/x_{i-1})}{\pi}\; z_i P_{ff}(z_i)
\theta_{x_{i-1}-x_i>\lambda e^{-t_i}},
    \qquad z_i=x_i/x_{i-1}.
\]
With the help of the following transformation of the integration variables
\[
  \int dx_i \frac{x_i}{x_{i-1}} 
  \frac{\theta_{x_{i-1}-x_i>\lambda e^{-t_i}}}{1-z_i}
= \int dx_i \frac{\theta_{x_{i-1}-x_i>\lambda e^{-t_i}}}{x_{i-1}-x_i}\; x_i
= \int\limits_{\lambda e^{-t_i}} \frac{dy_i}{y_i}\; x_i, \quad y_i\equiv x_{i-1}-x_i,
\]
we obtain
\begin{equation}
\begin{split}
&\frac{x}{u}
  G_{ff}(K^B;t_{b},t_{a},x,u)
  = e^{-\Phi_f(t_b,t_a|u)} 
    \delta_{x=u}+
\\&~
   +\sum_{n=1}^\infty \;
    \bigg[ \prod_{i=1}^n \;
      \int\limits^{t_b}_{t_{i-1}} dt_i\!\!\!
      \int\limits^1_{\lambda e^{-t_i}} \frac{dy_i}{y_i}\;
      \frac{\alpha_S(y_ie^{t_i})}{\pi} z_i(1-z_i)P_{ff} (z_i)
    \bigg]
   e^{-\sum_{j=0}^n\Phi_f(t_j,t_{j-1}|x_{j-1})}
   \delta_{x-u=\sum_{j=1}^n y_j}.
  \end{split}
\end{equation}
The simplification for the kernels and form-factor, to be compensated later on
by the MC weight, reads as follows:
\begin{equation}
\begin{split}
&\prod_{i=1}^n 
    \int\limits_{\lambda e^{-t_{i}}} \frac{dy_i}{y_i}\;
    \alpha_S(y_ie^{t_i})
    z_i (1-z_i) P_{ff} (z_i)
   e^{-\sum_{j=0}^n\Phi_f(t_j,t_{j-1}|x_{j-1})}
 \rightarrow
\\&~~~
 \rightarrow
    e^{-\mathbf{\Phi}_f(t_b,t_a|1-x)} 
    \prod_{i=1}^n \;
    \int\limits_{\lambda e^{-t_{i}}} \frac{dy_i}{y_i}\;
    \alpha_S(y_ie^{t_i}) A_{ff}=
\\&~~~
= e^{-\mathbf{\Phi}_f(t_b,t_a|1-x)} 
  \prod_{i=1}^n \;
    \int\limits_{\ln\lambda-t_{i}} dv_i\;
    \alpha_S(e^{t_i+v_i}) A_{ff}
= e^{-\mathbf{\Phi}_f(t_b,t_a|1-x)} 
  \prod_{i=1}^n \;
  \int\limits_{v_0(t_i)} dv_i\; \bar{\Pbbm}(t_i,v_i),
\\&
\bar{\Pbbm}(t_i,v_i) 
  = \alpha_S(e^{t_i+v_i})A_{ff},
\quad v_i=\ln(y_i),
\quad v_0(t)=\ln\lambda - t,
  \end{split}
\end{equation}
where $\mathbf{\Phi}_f$ is that of  eq.~(\ref{eq:PhiC}).
With  the above definitions we obtain
\begin{equation}
\begin{split}
&\frac{x}{u}
  G_{ff}(K^B;t_{b},t_{a},x,u)
  = e^{-\mathbf{\Phi}_f(t_b,t_a|1-x)}h(x,u)
  \bigg\{
  \delta_{\exp(v)=0}+
\\&~~~~~~~~~~~~~~~~~~
  +\sum_{n=1}^\infty \;
      \bigg[ \prod_{i=1}^n 
      \int\limits^{t_b}_{t_{i-1}} dt_i\;
      \int\limits_{v_0(t_i)}^{v} dv_i\; 
      \bar{\Pbbm}(t_i,v_i)
      \bigg]
   \delta_{\exp(v) =\sum_{j=1}^n \exp(v_j)}\;
   w^{\bar{\Pbbm}}
   \bigg\},
\end{split}
\end{equation}
where
\begin{equation}
v=\ln(u-x),\quad \Psi(v)=e^v, \quad \Psi'(v)=e^v=u-x
\end{equation}
and
\begin{equation}
\begin{split}
& w^{\bar{\Pbbm}}=
   e^{ \mathbf{\Phi}_f(t_b,t_a|u)-\sum_{j=0}^n\Phi_f(t_j,t_{j-1}|x_{j-1})}
   \prod_{i=1}^n \;
   \frac{\alpha_S(e^{t_i+v_i}) z_i(1-z_i)P_{ff} (z_i)}{\pi \bar{\Pbbm}(t_i,v_i)},
\\&
  h(x,u)=e^{\mathbf{\Phi}_f(t_b,t_a|1-x)-\mathbf{\Phi}_f(t_b,t_a|u)}.
\label{eq:wp-C}
\end{split}
\end{equation}
The factor%
\footnote{It will be efficiently dealt with
 by the general purpose MC tool FOAM~\cite{foam:2002}, see the next section.}
$h(x,u)$ is compensating for the deliberate use
of $\mathbf{\Phi}_f(t_b,t_a|u)$ in the weight $w^{\bar{\Pbbm}}$,
with the aim of ensuring  $w^{\bar{\Pbbm}}\leq 1$.
Since the simplified kernel $\bar{\Pbbm}$ is 
the same as in the previous case (B), the standardized
kernel and form-factor are also the same:
\begin{equation}
 R(v;t_b,t_a)
  =A_{ff}\frac{2}{b_0} \varrho_2(\tB_b+v,\tB_a+v;\tB_\lambda),\quad
  K(v) = \partial_v R(v;t_b,t_a),
\label{eq:Kv-C}
\end{equation}
Moreover, the upper phase space boundary is also the same $v_x=\ln(1-x)$, hence
\[
  R(v_x;t_b,t_a)=\mathbf{\Phi}_f(t_b,t_a|v_x).
\]
The standardized formula for the MC reads as follows
\begin{equation}
\begin{split}
&\frac{x}{u}
  G_{ff}(K^B;t_{b},t_{a},x,u)
= h(x,u)
   \bigg\{
   e^{-R(v_x)}\delta_{v=v_0(t_b)}
  +\theta_{v>v_0(t_b)}\frac{1}{\Psi'(v)}
   R'(v) e^{R(v)-R(v_x)}
\\&~~\times
   \sum_{n=1}^\infty
   P(n-1|R(v))
   \bigg[
      \prod_{i=1}^n  \int_{0}^{1}d\xi_i \int_0^1 d\sigma_i\;
   \bigg]
   \frac{\delta_{\max_j \xi_j=1}}{n}\;
   w^\#(\mathbf{\xi})\;
   w^{\mathbbm{P}}({\bf t}^{\Pmf_{\bf t}},{\bf v}^{\Pmf_{\bf t}})
   \bigg\}.
\end{split}
\end{equation}
The final integral, defining
the distribution to be implemented in the CMC, takes the following form:
\begin{equation}
\label{eq:GintegC}
\begin{split}
&\int_x^1 du
  G_{ff}(K^B;t_{b},t_{a},x,u)
= \int_0^1 dU\;
   \frac{u}{x}h(x,u)
   \bigg\{
   \theta_{U<\exp(-R(v_x))}|_{v=v_0(t_b)}
  +\theta_{U>\exp(-R(v_x))}
\\&~~~~~~~~\times
   \sum_{n=1}^\infty
   P(n-1|R(v))
   \bigg[
      \prod_{i=1}^n  \int_{0}^{1}d\xi_i \int_0^1 d\sigma_i\;
   \bigg]
   \frac{\delta_{\max_j \xi_j=1}}{n}\;
   w^\#(\mathbf{\xi})\;
   w^{\mathbbm{P}}({\bf t}^{\Pmf_{\bf t}},{\bf v}^{\Pmf_{\bf t}})
   \bigg\},
\end{split}
\end{equation}
where%
\footnote{In this case $R'(v)/\Psi'(v)=dR/du$.}
 $U=U(x,u)=e^{R(v)-R(v_x)}=e^{R(\ln(u-x))-R(\ln(1-x))}$.
The weight $w^\#$ is evaluated according to eq.~(\ref{eq:w-hash}).
All important differences with the previous cases (A) and (B)
are hidden in the definitions/constructions of the components
of the above formula.
The only explicit difference is in the presence of the factor $h(x,u)$.

\subsection{Summary on CMC for bremsstrahlung}

\begin{table}[!h]
\centering
\begin{tabular}{|c|c|c|c|c|c|c|c|c|}
\hline\hline
$X$ & $v$  & $v_i$ & $\Psi(v)$ &
$v_0$ & $\bar{\Pbbm}(t_i,v_i)$&
$ w^{\mathbbm{P}}$ & $K(v)$,\;$R(v)$
\\
\hline
$A$ &$1-\frac{x}{u}$& $1-\frac{x_i}{x_{i-1}}$& $\ln\big(1-e^v\big)$ &
$\ln\epsilon$ & $\alpha_S(e^{t_i})A_{ff}$        &
 Eq.(\ref{eq:wp-A}) & Eqs.(\ref{eq:Kv-A},\ref{eq:Rv-A})
\\
$B$ & $1-\frac{x}{u}$& $1-\frac{x_i}{x_{i-1}}$& $\ln\big(1-e^v\big)$ &
$\ln\lambda-t$& $\alpha_S(e^{t_i+v_i})A_{ff}$  &
 Eq.(\ref{eq:wp-B}) & Eqs.(\ref{eq:Kv-B},\ref{eq:Rv-B})
\\
$C$ & $u-x$        &           $x_{i-1}-x_i$& $e^v$ &
$\ln\lambda-t$& $\alpha_S(e^{t_i+v_i})A_{ff}$  &
 Eq.(\ref{eq:wp-C}) & Eq.(\ref{eq:Kv-C})
\\
\hline\hline
\end{tabular}
\caption{\sf
 For three types of the evolution kernel $\Keu^{(X)}$, $X=A,B,C$
 list of components in the generic eq.~(\ref{eq:Dgener2}) for CMC.
 Variable $v_x=\ln(1-x)$ is always the same.
}
\label{tab:components}
\end{table}
Before we extend our CMC to the complete evolution with quark gluon transitions,
let us summarize the case of pure bremsstrahlung.
As we have seen, all three cases of the kernels (A--C) are compatible
with the generic formula of eq.~(\ref{eq:Dgener2}),
provided we identify (define) properly all components there.
These components are collected and compared
in Table \ref{tab:components} for all three cases (A--C).
The appearance, in case C, of the extra factor $h(x,u)$ should be kept in mind.

\section{Complete CMC with quark gluon transitions}
\label{sec:completeQG}
Monte Carlo simulation/integration of variables related to
quark gluon transitions is managed by the
general purpose MC tool FOAM~\cite{foam:2002}.
It has to be provided with the user-defined integrand, the so-called
{\em density function}. Arguments of this function
have to be inside a unit hyperrectangle (or a simplex).
Starting from eq.~(\ref{eq:SoluX}) we are going to reorganize
explicitly its integration variables,
see also the scheme in Fig.~\ref{fig:hierarchX},
paying attention to the integration order:
\begin{equation}
\label{eq:Q-G-raw}
\begin{split}
&D_f(t,x)=
     \int\limits_x^1 dx_0\; 
     G_{ff}(K^B;t, t_{0};x,x_0)\;
     D_f(t_0,x_0)+
\\&~~~
  +\sum_{n=1}^\infty \;
   \sum_{f_{n-1},\dots,f_{1},f_{0}}
    \Bigg[ \prod_{k=1}^n\;
     \int\limits_{t_{k-1}}^t dt_k\;
     \Bigg]\;
    \int\limits_{x}^1 dx_n\;
   G_{ff}(K^B;t,t_n,x,x_n)\;
\\&~~~~~~\times
   \Bigg[ \prod_{k=1}^n \;
     \int\limits_{x_k}^1 dx'_k\;
     \Peu^A_{f_kf_{k-1}}(t_k,x_k,x'_k)\;
     \int\limits_{x'_k}^1 dx_{k-1}\;
     G_{f_{k-1}f_{k-1}}(K^B;t_k,t_{k-1},x'_k,x_{k-1}) \Bigg]\;
      D_{f_0}(t_0,x_0),
\end{split}
\end{equation}
where $f_n\equiv f$.
Here, and in the following, we understand that the following
integration order for all multiple integrations%
\footnote{ This is the same order as for the operator products
   and in the time-ordered exponentials.} is used.
\[
  \prod_{i=1} \int d\alpha_i= \int d\alpha_n \int d\alpha_{n-1}
     \dots \int d\alpha_2 \int d\alpha_1.
\]
Let us now re-parametrize the above integral into a form better suited
for integration by FOAM, that is in terms of $3n+1$ variables:
$U_0\in(0,1)$ and $U_k ,\alpha_k,\beta_k\in(0,1), k=1,2,...n$:
\begin{equation}
\label{eq:Q-G-foam}
\begin{split}
&D_f(t,x)=
     \int\limits_0^1 dU_0(x,x_0)\; 
     H^{(X)}(x_0,x)
     \int_{G(t_{b},t_{a})} dG^0\;W^G_0
     D_f(t_0,x_0)+
\\&~~~
  +\sum_{n=1}^N \;
   \sum_{f_{n-1},\dots,f_{1},f_{0}}
    \Bigg[ \prod_{k=1}^n\;
     \int\limits_0^1 (t-t_{k-1})d\alpha_k\;
     \Bigg]\;
    \int\limits_{0}^1 dU_n(x,x_n)\;
     H^{(X)}(x_n,x)
    \int_{G(t_b,t_{n-1})} dG^{(n)}\; W^G_n\;
\\&~~~~~~\times
   \Bigg[ \prod_{k=1}^n \;
     \int\limits_0^1 d\beta_k\; (1-x_k)
     \Keu^A_{f_kf_{k-1}}(t_k,x_k,x'_k(\beta_k))\;
\\&~~~~~~~~~~~\times
     \int\limits_0^1 dU_{k-1}(x'_k,x_{k-1})\;
     H^{(X)}(x_{k-1},x'_k)
     \int_{G(t_k,t_{k-1})} dG^{(k-1)}\; W^G_{k-1}
   \Bigg]\;
      D_{f_0}(t_0,x_0),
\end{split}
\end{equation}
where for all three cases $X=A,B,C$,
the multi-differentials $dG$ are used%
\footnote{See eqs.~(\ref{eq:GintegA}) and (\ref{eq:GintegC}).}

\begin{equation}
\int_x^1 du\;
  G_{ff}(K^B;t_{b},t_{a},x,u)
= \int_0^1 dU\;
  H^{(X)}(x,u)\;
  \int_{G(t_b,t_a)} dG\; W^G,
\end{equation}
they  are defined in the following way%
\footnote{The index $k$ in $dG^{(k)}$ reminds us of the type of the
  parton type $f_k$ used implicitly in the distribution.
}
\begin{equation}
\label{eq:Ginteg}
\begin{split}
\\&
  \int_{G(t_b,t_a)} dG^{(k)} \equiv
   \theta_{U<\exp(-R(v_x))}|_{v=v_0(t_b)}+
\\&~~~~~~~~
  +\theta_{U>\exp(-R(v_x))}
   \sum_{n=1}^\infty
   P(n-1|R(v))
   \bigg[
      \prod_{i=1}^n  \int_{0}^{1}d\xi_i \int_0^1 d\sigma_i\;
   \bigg]
   \frac{\delta_{\max_j \xi_j}=1}{n},
\\&
  \int\limits_0^1dU\; \int_{G(t_b,t_a)} dG \equiv 1.
\end{split}
\end{equation}
As in case  of $dG^{(k)}$, the   weight 
\begin{equation}
\label{eq:Gintega}
   W^G=
   w^\#(\mathbf{\xi})\;
   w^{\mathbbm{P}}({\bf t}^{\Pmf_{\bf t}},{\bf v}^{\Pmf_{\bf t}})
\end{equation}
 depend on the $t$ range 
(in the above definition $t_b$, $t_a$ is used). 
For each occurence of $dG$ and $W^G$ in (\ref{eq:Q-G-foam}), 
different $t$ range  is used, as is  seen from 
the context. We additionally  mark the difference 
for the later use.

For our three example kernels we have
\begin{equation}
\begin{split}
&H^{(A)}(x,u)=H^{(B)}(x,u)=\left(\frac{u}{x}\right)^2,\\
&H^{(C)}(x,u)=\frac{u}{x}\; h(x,u).
\end{split}
\end{equation}
The summation over the number of quark gluon transitions
$n$ is also mapped into one of the FOAM variables,
after truncation to a finite $N$.
In fact, the precision level of $\sim 10^{-4}$ is achieved for $N=4$,
see numerical results below.
Following the prescription of ref.~\cite{Jadach:2005bf}
the sum $\sum_{f_{n-1},\dots,f_{1},f_{0}}$ can be reduced just to a single term.
Also, as in ref.~\cite{Jadach:2005bf}, the additional mapping to the
$\tau_i=\tau(t_i)$ variables is done 
in order to compensate partly for the $t$-dependence of $\alpha_S(t_i)$
in the $\Peu$ kernels. The purpose of this mapping is to  boost
slightly the efficiency of FOAM.

FOAM generates all $3N+2$ variables (including $n$)
according to the integrand of eq.~(\ref{eq:Q-G-foam}),
omitting from it the following MC weight
\begin{equation}
\label{eq:MCweight}
W = \prod_{i=0}^n W^G_i.
\end{equation}
Let us stress, that the calculation of this weight can be  completed only 
after gluons are generated
for all $n+1$ bremsstrahlung segments first.
Also, as in ref.~\cite{Jadach:2005bf}, FOAM treats distributions
as continuous in $U_i$-variables, ignoring $\delta$-like
structure in $x'_k-x_{k-1}$, corresponding to the no-emission case.
This is very important and powerful method.
The  $\delta$-like  no-emission terms are replaced by
the integrals over finite intervals in $U_i$,
exactly as in eq.~(\ref{eq;Uintegral}).

\begin{figure}[!ht]
  \centering
  {\epsfig{file=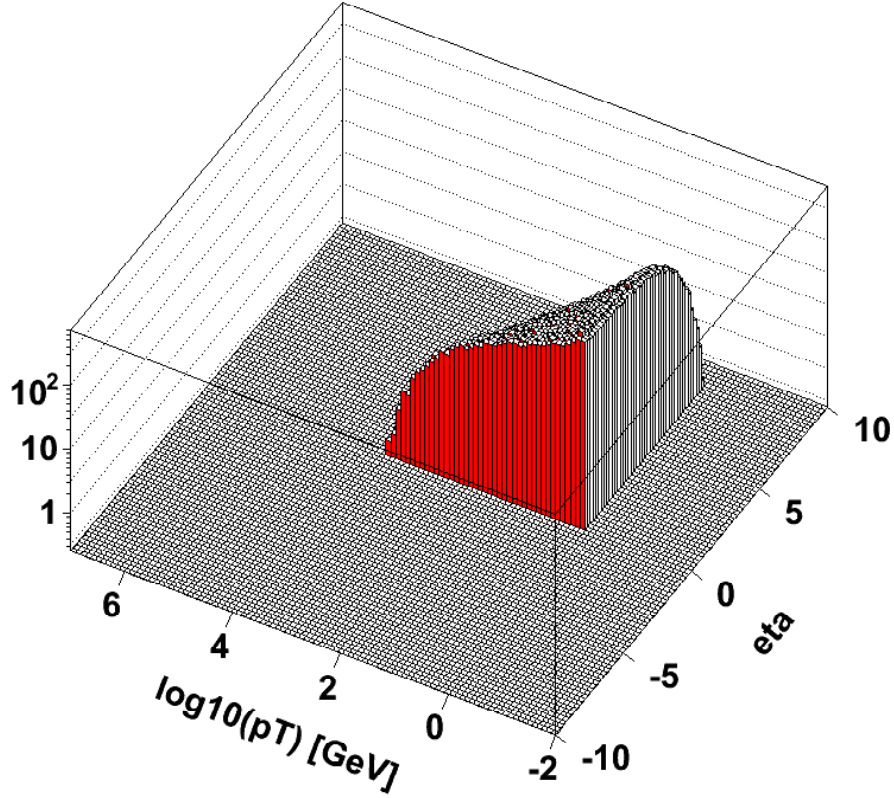, width=120mm}}
  \caption{\sf
    The distribution of rapidity and log of $k^T$ from CMC for $2E_h=1000$~GeV,
    $\lambda=1$~GeV and the kernel type (C); pure bremsstrahlung case.
    }
  \label{fig:CMIkTrap}
\end{figure}

With all the above formalism at hand, we can formulate our
CMC algorithm similarly as in the general LL DGLAP case:
\begin{itemize}
\item
  Generate super-level variables $n$, $t_i$ $x'_i$ and $x_i$
  with the help of the general-purpose MC tool FOAM according to 
  eq.~(\ref{eq:Q-G-foam}), and neglecting $W=\prod W^G_i$.
  The parton types $f_i, i=0,1,2,3...,n-1$ are determined from $f=f_n$
  according to prescription of ref.~\cite{Jadach:2005bf}.
\item
  In the above, the number of flavour transitions 
  ($G\to Q$ and $Q\to G$) is limited
  to $n=0,1,2,3,4$, aiming at the precision of $\sim 0.2\%$.
\item
  For each $i$-th pure gluon bremsstrahlung segment
  the emissions variables are generated using
  the dedicated bremsstrahlung CMC of section~\ref{sec:gluonstrahlung},
  according to eq.~(\ref{eq:Ginteg}). The weights $W^G_i$ are calculated.
\item
  Generated MC events are weighted with $W=\prod W^G_i$. They are optionally
  turned into weight=1 events using the conventional rejection method.
\item
  The four-momenta $k_i^\mu$ and $q_i^\mu$ are reconstructed 
  out of evolution variables and azimuthal angles%
\footnote{Azimuthal angles are generated uniformly.}.
 \end{itemize}
The above algorithm is already implemented in the form of a 
program in C++ and tested using upgraded version of the Markovian MC
of refs.~\cite{Jadach:2003bu} and \cite{Golec-Biernat:2006xw}.
The numerical results are documented in the following section.

\begin{figure}[!ht]
  \centering
  {\epsfig{file=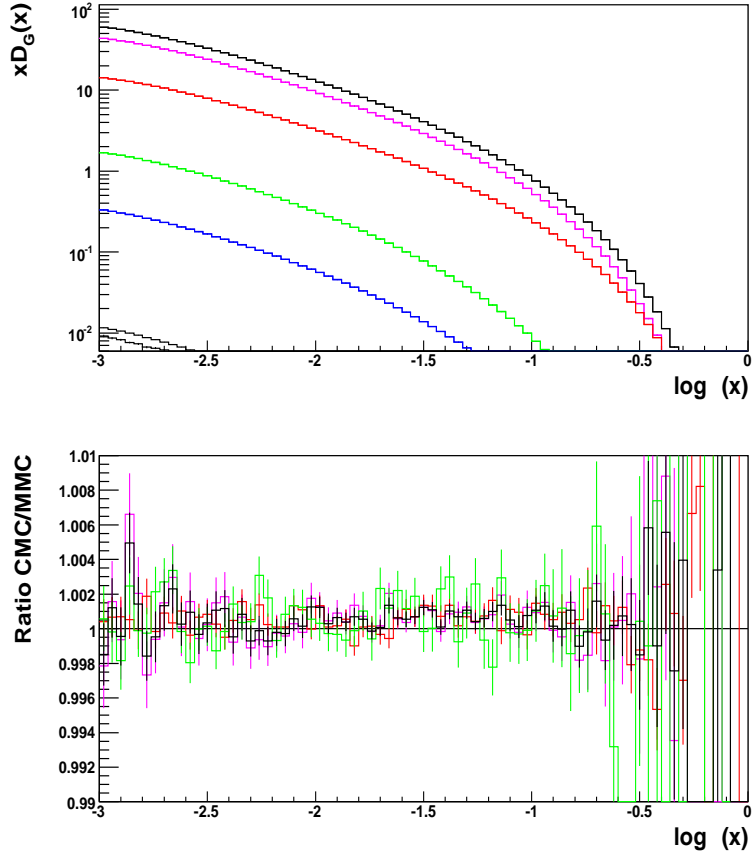, width=110mm, height=120mm}}
  \caption{\sf
    The gluon distribution from CMC and MMC for evolution with
    the kernel (C'), $\lambda=1$~GeV and $e^{t_{\max}}=2E_h=1$~TeV.
    Contributions from fixed number of the quark gluon transitions
    $n=0,1,2,3,4$ are also shown.
    The ratios CMC/MMC in the lower plot are given separately for the total result
    and for the number of quark gluon transition $n=0,1,2$.
    The MC statistics is $4.5\cdot 10^{9}$ weighted
    events for CMC and $10^{10}$ for MMC.
    }
  \label{fig:CMCIfdlnxG}
\end{figure}

\begin{figure}[!ht]
  \centering
  {\epsfig{file=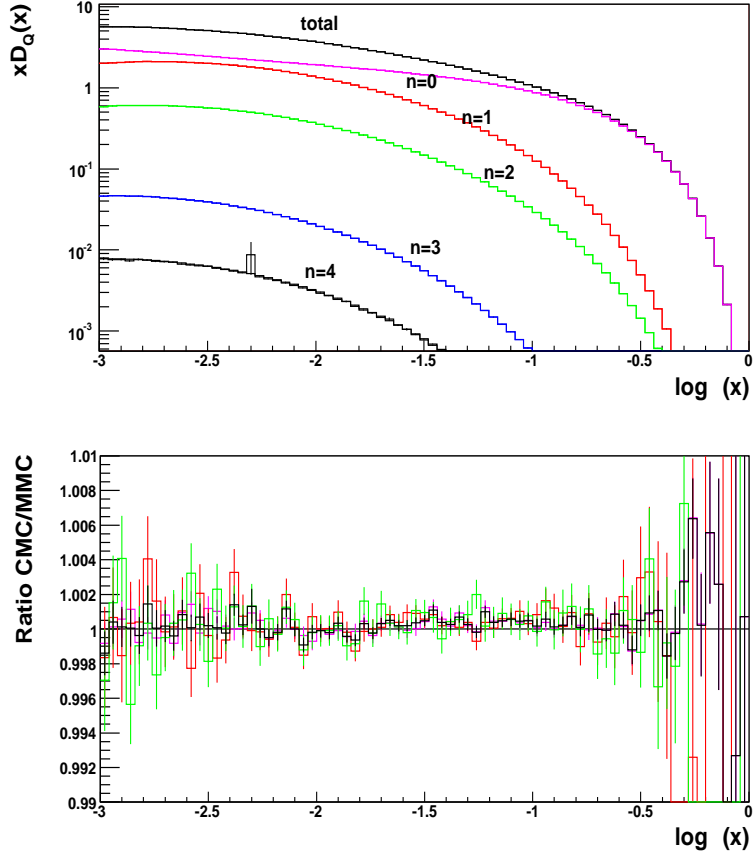, width=110mm, height=120mm}}
  \caption{\sf
    The Quark distribution from CMC and MMC for evolution with
    the kernel (C'), $\lambda=1$~GeV and $e^{t_{\max}}=2E_h=1$~TeV.
    Contributions from fixed number of the quark gluon transitions
    $n=0,1,2,3,4$ are also shown.
    The ratios CMC/MMC in the lower plot are given separately for the total result
    and for the number of quark gluon transition $n=0,1,2$.
    The MC statistics is $10^{10}$ weighted
    events for both CMC and MMC.
    }
  \label{fig:CMCIfdlnxQ}
\end{figure}

\section{Numerical tests}
Most of the numerical tests proving that the new CMCs
of this paper work correctly were done by means of comparison with
the updated version of the Markovian MC program MMC \cite{SingleMMC},
which is a descendant of that described in
refs.~\cite{Golec-Biernat:2006xw,Jadach:2003bu}.
The version of MMC used below implements exactly the same type evolution
with exactly the same kernels and boundary conditions.
We can therefore expect numerical agreement of CMC and MMC to
within statistical MC error, which will be of the order of $10^{-3}$,
for about $10^{10}$ MC events
(and for about $10^2$ different values of $x$ in a single MC run).

We start, however, with the simple tests in which we verify
the correctness of the mapping of the evolution variables into four-momenta.
In Fig.~\ref{fig:CMIkTrap} 
the distribution of rapidity and $k^T_i$ of the emitted gluons is shown.
Sharp cut-offs corresponding to
the minimum rapidity $\eta\geqslant 0$ (maximum $t$) and 
minimum $k^T\geqslant 1$~GeV are clearly visible in the plot.
Note that this plot shows the same triangular area of the logarithmic Sudakov plane as
in Fig.~\ref{fig:sudak0}(a), but now populated with the MC events.

\begin{figure}[!ht]
  \centering
  {\epsfig{file=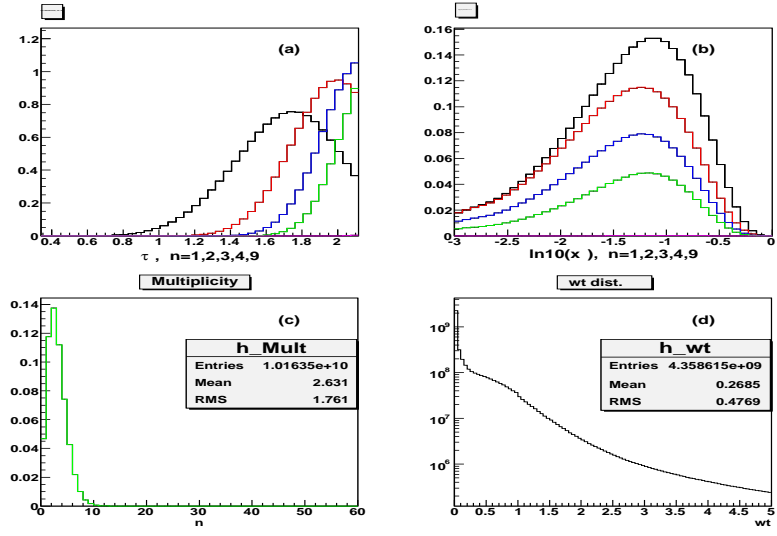, width=110mm, height=72mm}}
  \caption{\sf
    The distributions of $\tau_i$ and $x_i$ and the emission multiplicity from
    the CMC and MMC programs.
    The results are from the MC run for gluon in proton.
    }
  \label{fig:CMCIfexclG}
\end{figure}

\begin{figure}[!ht]
  \centering
  {\epsfig{file=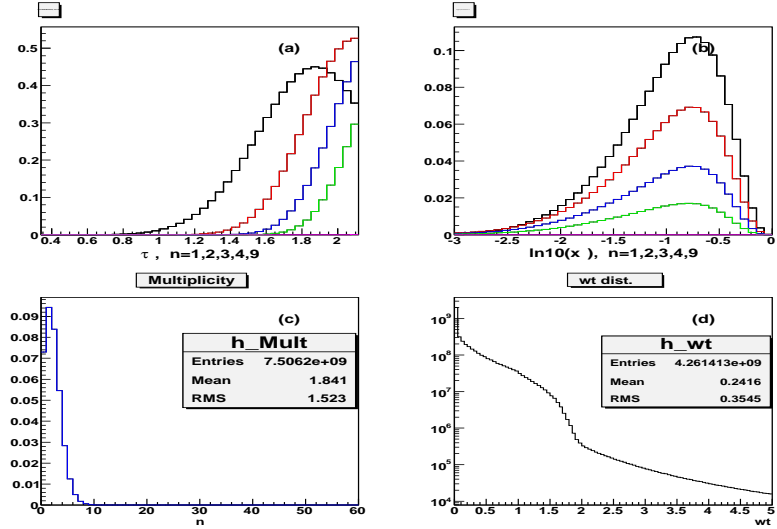, width=110mm,  height=72mm}}
  \caption{\sf
    The distributions of $\tau_i$ and $x_i$ and the emission multiplicity from
    the CMC and MMC programs.
    The results are from the MC run for quark in proton.
    }
  \label{fig:CMCIfexclQ}
\end{figure}

\subsection{Testing CMC versus MMC}

We examine results of the evolution from the initial energy
scale $\lambda=1$ GeV up to final energy scale of $E_h=1$~TeV,
using exactly the same initial quark and gluon distributions
in a proton at the
$Q_0=e^{t_0}=\lambda=1$~GeV scale as in ref.~\cite{Jadach:2005bf}
(they were also used in ref.~\cite{Jadach:2003bu}).

In Fig.~\ref{fig:CMCIfdlnxG} the distributions of gluons at $E_h=1$~TeV
are shown, while in Fig.~\ref{fig:CMCIfdlnxQ}
the corresponding results for quarks,
$Q=q+\bar q$, are exposed, also at the same high energy scale $E_h=1$\,TeV.
Numerical results are provided for the evolution type (C')
(see section \ref{sec:kernels}), that is for $\alpha_S(k^T)$ and the evolution
time being identical with the rapidity of the emitted particle.
In these two figures we compare the gluon and quark distributions obtained
from the current CMC program and from the updated MMC program \cite{SingleMMC}.
The principal numerical results from both programs,
marked as ``total'' in the upper plot of both figures,
are indistinguishable.
We therefore plot their ratio in the lower plot of both figures.
The parton distributions from the CMC and MMC programs {\em agree perfectly},
within the statistical errors of $\sim 10^{-3}$
in the entire range of $x$.
In the same plots we include individual contributions from a
fixed number of the quark gluon transitions $n=0,1,2,3,4$,
and their ratios.
Again very good agreement between CMC and MMC results is seen%
\footnote{Certain disagreements for $n=4$, which affect total
   result at the $10^{-4}$ level can be traced back to
   well understood inefficiencies of FOAM at higher dimensions.}.

It should be stressed that the MMC program has been tested numerically at the
same $\sim 10^{-3}$ precision level for all types of the kernels (A--C)
by comparison with the semi-analytical (non-MC) program {\tt APCheb}~\cite{APCheb33}.
In addition, for the DGLAP case, MMC was also successfully compared with another
non-MC program {\tt QCDNum16}~\cite{qcdnum16}.
These dedicated tests of MMC using non-MC programs will be published separately
in the forthcoming paper~\cite{SingleMMC}.

Additionally, in Figs.~\ref{fig:CMCIfexclG} and \ref{fig:CMCIfexclQ}, 
we show comparisons of selected (semi-)exclusive distributions from CMC and MMC,
i.e. we have chosen for the test the distributions of the consecutive
evolution time variables $\tau_i(t_i)$, the energy fractions $x_i$
and the total parton emission multiplicity from both CMC and MMC.
The above distributions resulting from separate CMC and MMC runs are interposed --
no visible difference is seen within the resolution of the plots.
The above plots are for a final parton being a gluon
in Fig.~\ref{fig:CMCIfexclG} and a quark in Fig.~\ref{fig:CMCIfexclQ}.
In these plots one is testing a nontrivial aspect of the CMC
algorithm related to removing and restoring the $t$-ordering
described in section \ref{sec:torder}.
In fact, any departure from the {\em pairwise} ordering procedure
described in section \ref{sec:torder} would ruin the agreement
of CMC and MMC
seen in Figs.~\ref{fig:CMCIfexclG} and \ref{fig:CMCIfexclQ}!
The weight distribution from CMC is also shown in
the lower part of these figures.

All the above tests were done for the most interesting
case of CMC with the kernel (C).
What about numerical tests for cases (A) and (B)?
In ref.\cite{Jadach:2005bf}
numerical agreement within $\sim 10^{-3}$ statistical error between
CMC and MMC for the LL DGLAP case (A) was already documented.
While preparing this work, we have compared CMC and MMC
for case (B), $\alpha_S(e^{-t}\lambda)$,
obtaining equally good numerical agreement.
The corresponding plots%
\footnote{They were shown in the contribution
  by S.~Jadach to the Ustro\'n Conference, September 2005,
  see {\tt http://home.cern.ch/jadach/public/ustron05.pdf}.}
look quite similar to these in
Fig.~\ref{fig:CMCIfdlnxG} and Fig.~\ref{fig:CMCIfdlnxQ}.

\section{Discussions and summary}

We have generalized the constrained Monte Carlo algorithm
of ref.~\cite{Jadach:2005bf} from DGLAP
to two other more complicated types of the evolution equations,
one of them fully compatible%
\footnote{We did not include into the discussion the so called
 non-Sudakov form-factor for CCFM, case (C).
 However,  it is already included in the code.}
with the CCFM evolution equation \cite{CCFM}.
The above extentions of the CMC algorithm are implemented in the
computer program CMC, tested by comparisons it with the Markovian MC,
which also solves exactly the same evolution equations.

Let us point out that the most complicated version (C) of the evolution
equation elaborated in this work
follows closely the CCFM model as formulated in \cite{Marchesini:1995wr}
(except for the temporarily omitted non-Sudakov form-factors).
Its Markovian MC implementation {\tt SMALLX} was worked out
in refs.~\cite{Marchesini:1990zy,Marchesini:1991zy},
and later on exploited in the construction of the {\tt CASCADE} MC
of ref.~\cite{Jung:2000hk},
which is based on the backward evolution algorithm \cite{Sjostrand:1985xi}.
Our CMC program was tested against our own Markovian MC, see previous section.
However, it would be interesting to compare it also with
the above {\tt SMALLX} and {\tt CASCADE} programs.
We hope this will be done in the near future.

Since the CMC program of this work will be used as a building block
in the MC event generator for the Drell--Yan type processes and deep
inelastic lepton-hadron scattering, the explicit mapping of
the evolution variables into four-momenta of the emitted particles
is also defined, implemented and tested.
The evolution time is mapped into the rapidity variable (angular ordering).
A sharp cut-off is imposed on the $k^T$ and rapidity of the emitted particle.
The sharp cut-off in the rapidity will be useful when combining two CMCs
for two colliding initial state hadrons, with neither gaps nor overlaps
in the emission phase space.

The CMC algorithm was worked out in detail and tested
for three types of the evolution kernels and the phase--space limits.
It was purposely defined/described in such a way that it can be easily
generalized to other types of the evolution kernels.
In the following works it will be used as a component
in the MC modelling of the initial state parton shower
for showering of a single hadron in a more complete MC project for LHC.

\vspace{10mm}
\noindent
{\bf\Large Acknowledgments}
\vspace{2mm}\\
We would like to thank K. Golec-Biernat for the useful help and discussions.
We acknowledge the warm hospitality of the CERN Physics Department, where part
of this work was done.

\vfill\newpage

\appendix
\noindent
{\bf\Large Appendix}

\section{Auxiliary functions in form-factors}
\subsection{Triangle function $\varrho$}
\label{append:triangle}

\begin{figure}[!ht]
  \centering
  \epsfig{file=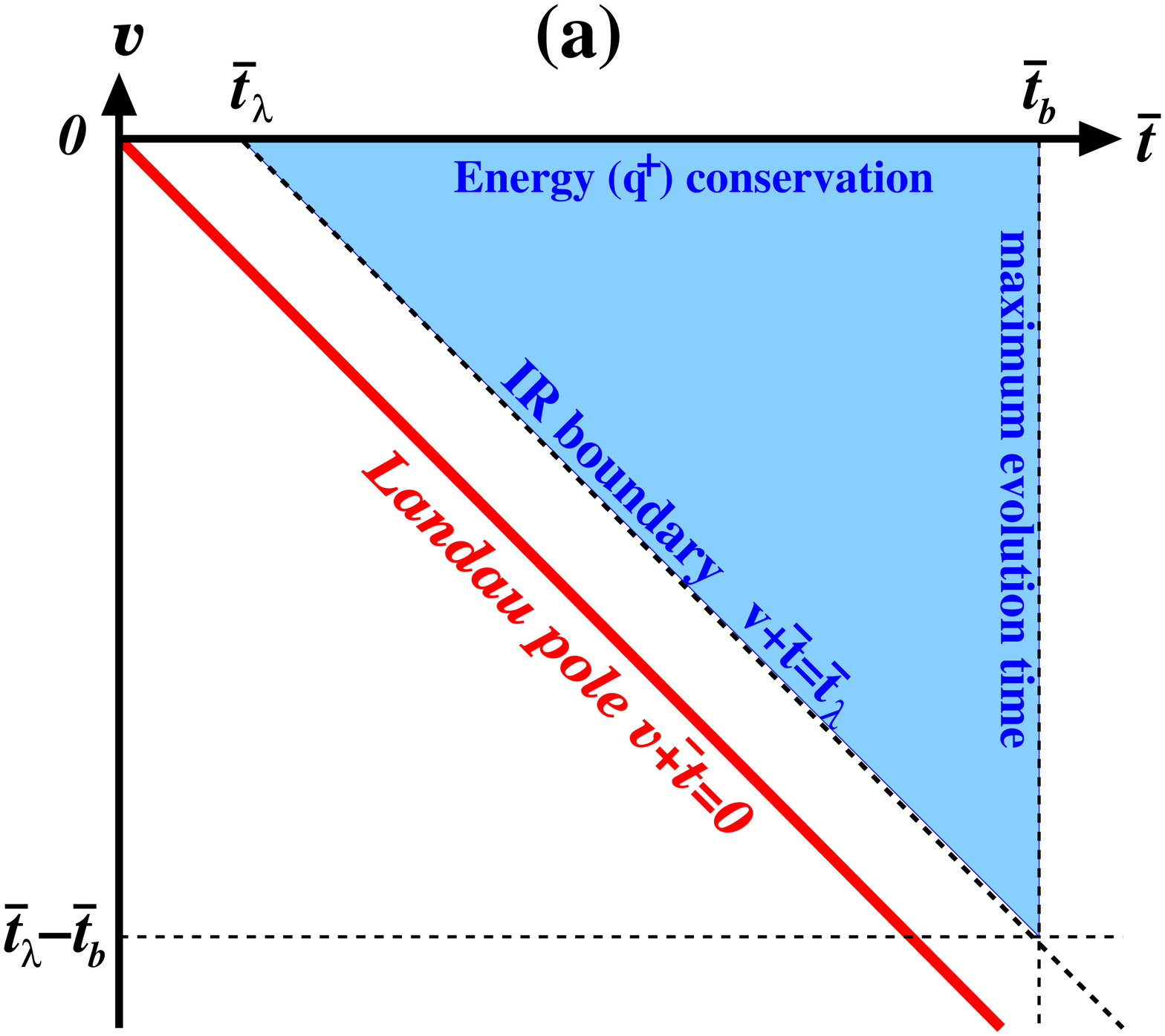, width=77mm}
  \epsfig{file=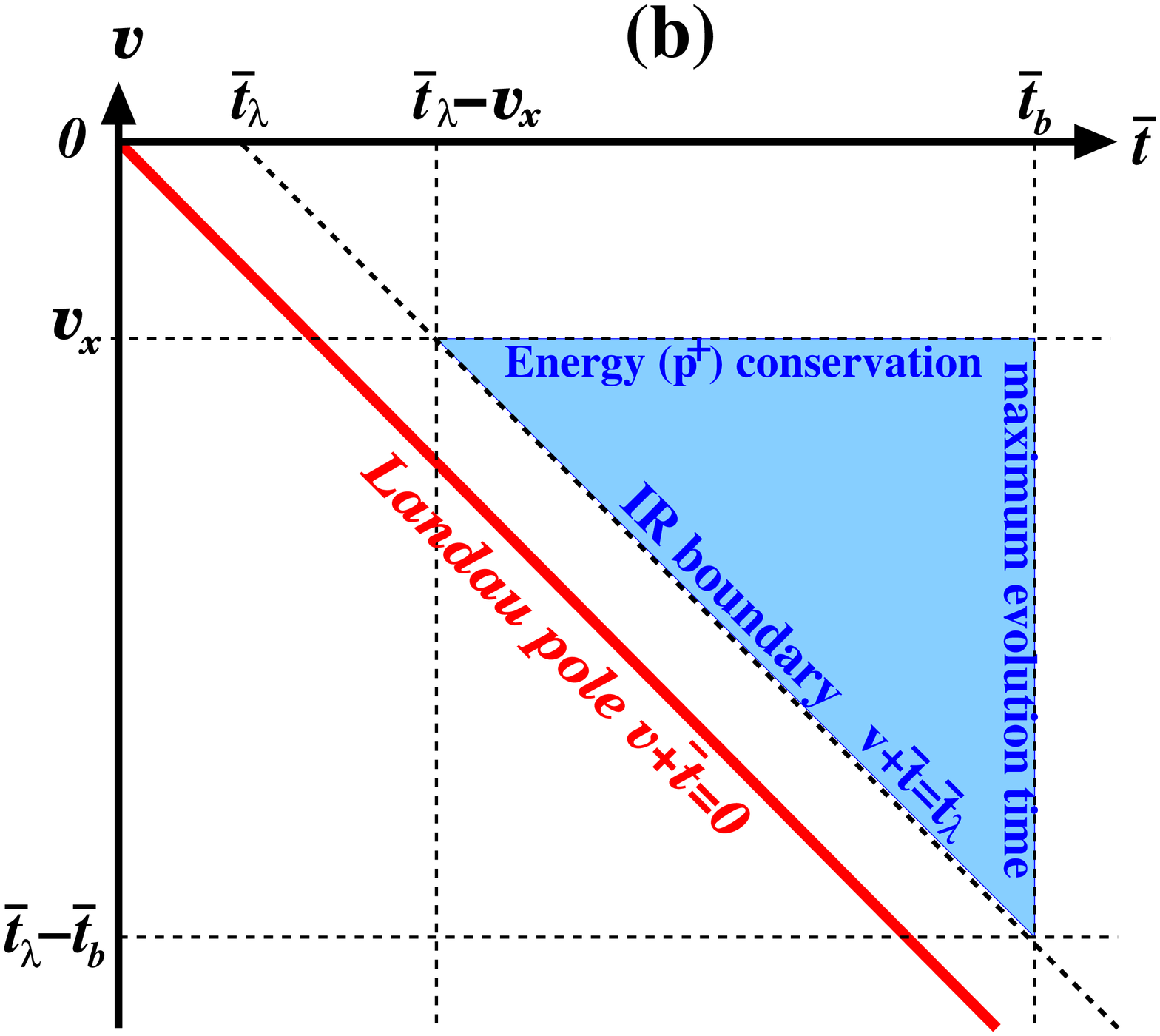, width=75mm}
  \caption{\sf
   The integration area in the definition of the functions:
   (a) $\varrho(\tB_b;\tBl)$ and (b) $\varrho(\tB_b +v_x;\tBl)$.
    }
  \label{fig:sudak-rho1}
\end{figure}
The simplest component function in the Sudakov formactor
in cases (B) and (C) of $z$-dependent $\alpha_S$ reads as follows
\begin{equation}
\varrho(\tB_b;\tBl) = \int^{\tB_b} d \tB \int^{0} d v \;
    \frac{\theta_{\tB + v > t_{\lambda}}}{\tB + v}
=\theta_{\tB_b>\tB_\lambda}
 \{\tB_b\; [\ln \tB_b -\ln\tB_\lambda - 1] + \tB_\lambda\},
\label{eq:rho}
\end{equation}
where $\lambda$ defines the IR cut-off.
Here and in the following, we follow certain notation rules
allowing us to write the above and similar functions in a compact way:
\begin{enumerate}
\item 
For all variables like $t$, $t_b$ and $t_\lambda$ bar over them means
$\tB=t-\ln\Lambda_0$, $\tB_b=t_b-\ln\Lambda_0$, $\tB_\lambda=t_\lambda-\ln\Lambda_0$, etc.,
where $\Lambda_0$ is that in eq.~(\ref{eq:alpha}),
i.e. the position of the Landau pole.
\item
Occasionally we shall omit the explicit dependence on $t_\lambda=\ln\lambda$,
that is we always understand
$\varrho(\tB_b) \equiv \varrho(\tB_b;\tBl)$.
\item
We always understand that $\int_a^b dx\equiv0$ when $b\leq a$.
\end{enumerate}
The area of the integration in eq.~(\ref{eq:rho}) is the triangle, depicted
if Fig.~\ref{fig:sudak-rho1}~(a).

The following similar integral, with the same integrand, but
slightly different triangular integration area,
depicted in Fig.~\ref{fig:sudak-rho1}~(b),
can be expressed using the same function $\varrho$:
\begin{equation}
\int^{\tB_b} d \tB \int^{v_x} d v 
    \frac{\theta_{\tB + v > \bar t_{\lambda}}}{\tB + v}
=\varrho(\tB_b+v_x;\tBl).
\label{eq:rho-vx}
\end{equation}

\subsection{Trapezoid function $\varrho_2$}
\label{append:trapez}
\begin{figure}[!ht]
  \centering
  \epsfig{file=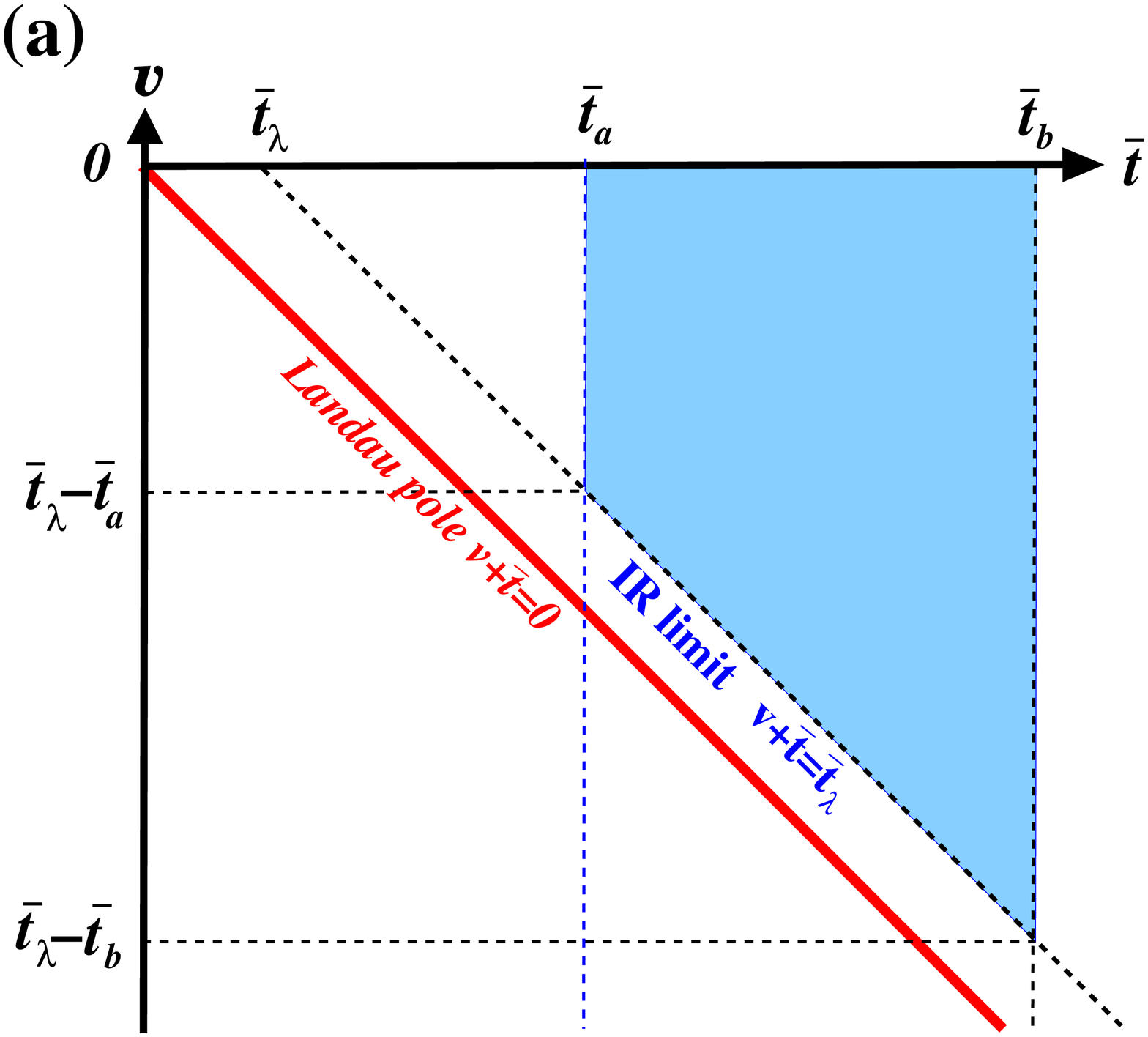, width=75mm}
  \epsfig{file=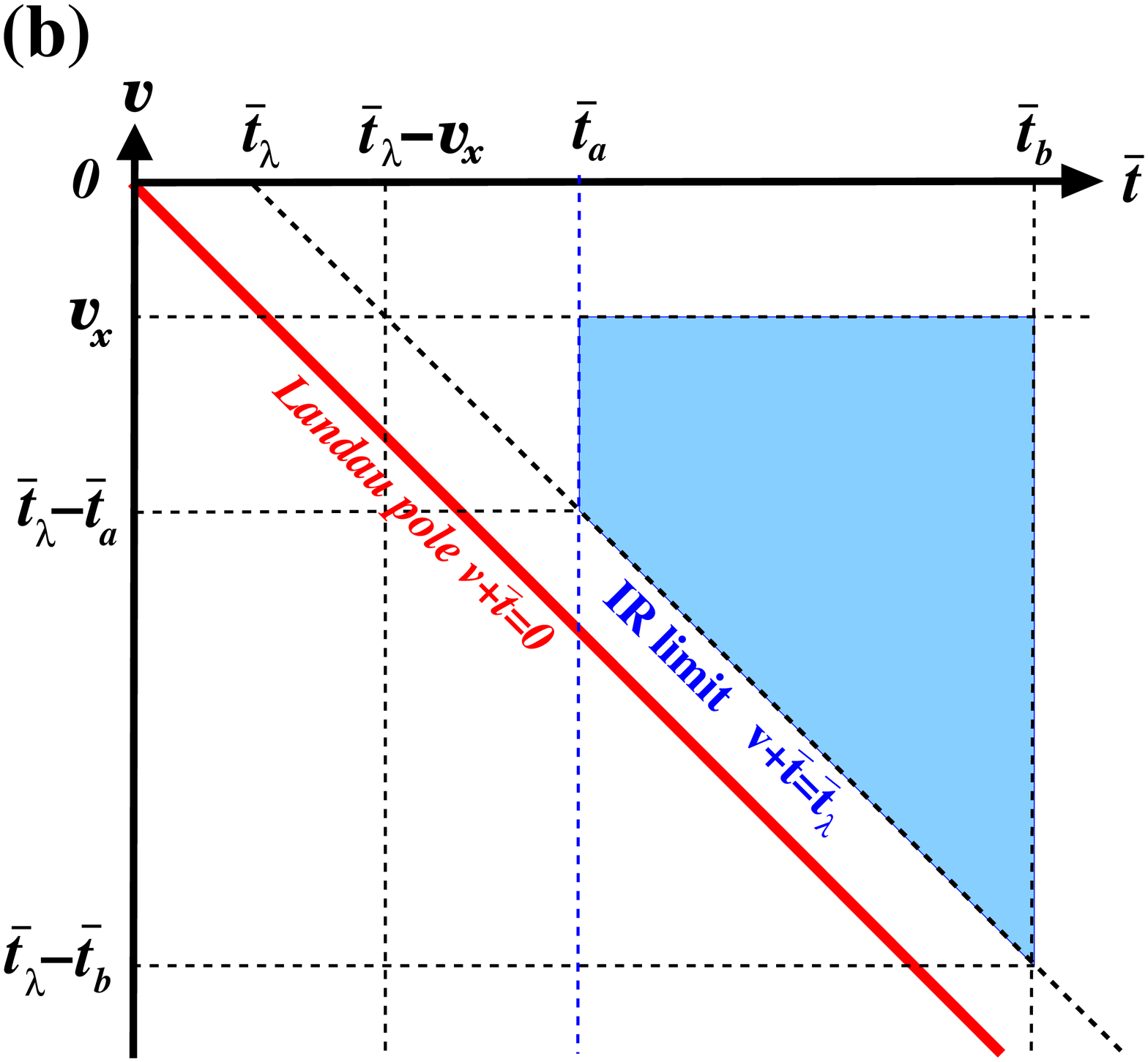, width=75mm}
  \caption{\sf
   The integration area in the functions:
   (a) $\varrho_2(\tB_a,\tB_b;\tBl)$ and 
   (b) $\varrho_2(\tB_a+v_x,\tB_b +v_x;\tBl)$.
    }
  \label{fig:sudak-rho2}
\end{figure}
The other basic IR-singular component function in the Sudakov form-factor reads
\begin{equation}
\varrho_2(\tB_a,\tB_b;\tBl) = \int^{\tB_b}_{\tB_a} d \tB \int^{0} d v 
    \frac{\theta_{\tB + v > \bar t_{\lambda}}}{\tB + v}
=\theta_{\tB_b>\tB_a}
 \{\varrho(\tB_b;\tBl)-\varrho(\tB_a;\tBl)\}
\label{eq:rho2}
\end{equation}
The corresponding integration area is the trapezoid
depicted in Fig.~\ref{fig:sudak-rho2}~(a).
The above calculation result is obvious if one notices that the trapezoid is
the difference of two overlapping triangles.
Similarly, the following integral with the trapezoid integration area
of Fig.~\ref{fig:sudak-rho2}~(a) is again expressed in terms of the functions
already defined above
\begin{equation}
\begin{split}
\varrho^{[1]}_2(v_x,\tB_a,\tB_b;\tBl) 
&=\int^{\tB_b}_{\tB_a} d \tB \int^{v_x} d v 
    \frac{\theta_{\tB + v > \bar t_{\lambda}}}{\tB + v}
\\&
 =\theta_{\tB_b>\tB_a}
\{\varrho(\tB_a+v_x;\tBl) -\varrho(\tB_b+v_x;\tBl)\}
=\varrho_2(\tB_a+v_x,\tB_b+v_x;\tBl).
\label{eq:rho2d}
\end{split}
\end{equation}
Let us stress that keeping $\theta_{\tB_b>\tB_\lambda}$
in the basic function $\varrho$ of eq.~(\ref{eq:rho}) is essential
for validity of the evaluation of all the following related functions and integrals.

\subsection{Non-singular functions and mapping}
\label{append:mapping}

\begin{figure}[!ht]
  \centering
  \epsfig{file=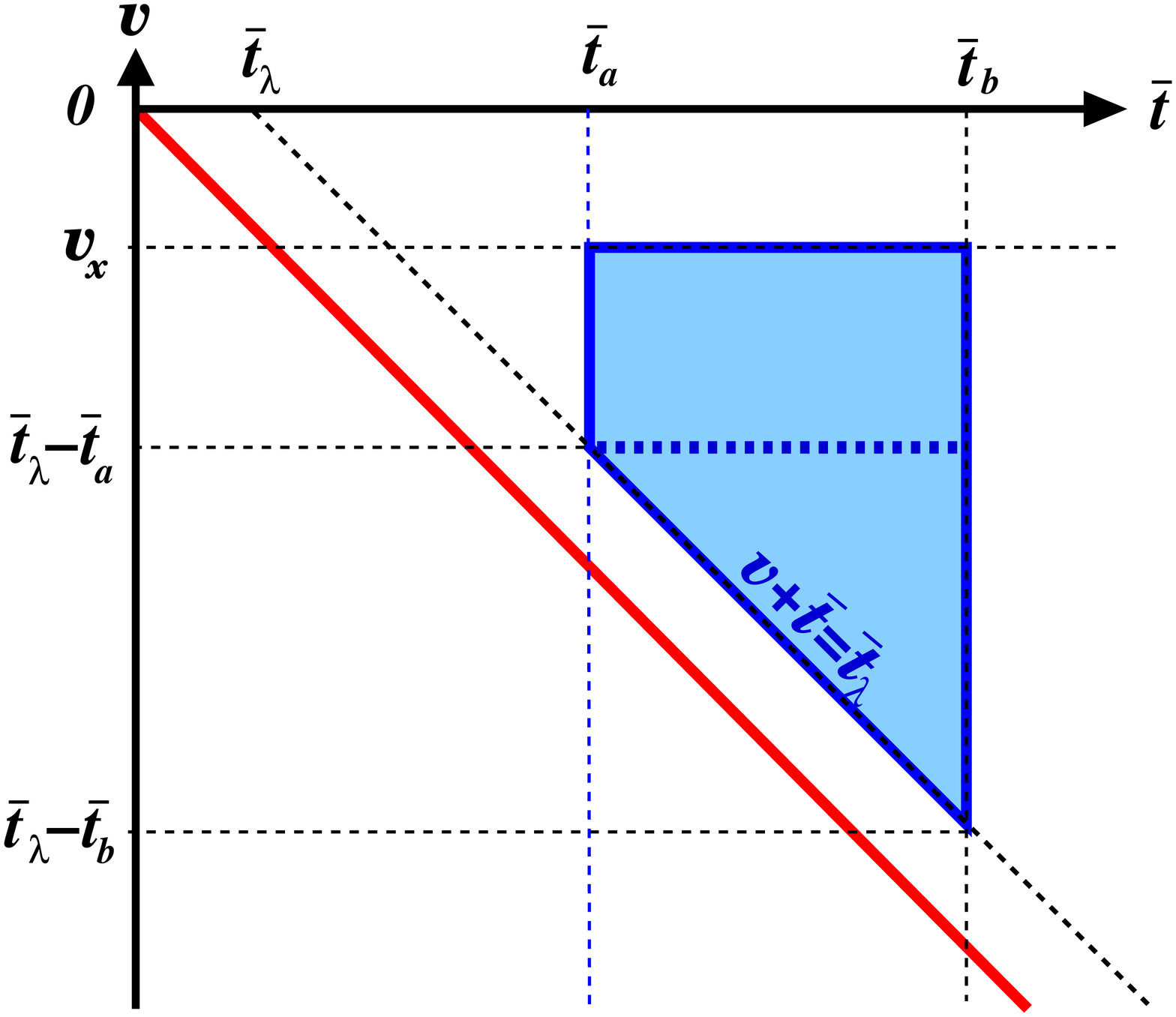, width=80mm}
  \caption{\sf
   The integration area in the function
   $\varrho^{[F]}_2(\tB_a,\tB_b,v_x;\tBl)$.
    }
  \label{fig:sudak-rho2F}
\end{figure}

Let us evaluate now the following integral, which is similar to $\varrho_2$,
except that we insert into the integrand the additional function $F(v,t)$:
\begin{equation}
\varrho^{[F]}_2(v_x,\tB_a,\tB_b;\tBl) 
  = \int^{\tB_b}_{\tB_a} d \tB \int^{v_x} d v \;
    \frac{\theta_{\tB + v > \bar t_{\lambda}}}{\tB + v} F(v,\tB).
\label{eq:rho2F}
\end{equation}
Of course, for $F=1$ we recover $\varrho_2$.
However, even in this case the following evaluation of $\varrho^{[F]}_2$ will be
interesting, because we shall swap the integration order and introduce
variable mapping, exactly the same as we have used for finding out $K(v)$
and eliminating $K(v)$ through the additional mapping $r(v)$.
Such a swapping integration order is also used for evaluation of
the non-IR part of the bremsstrahlung integral and flavour changing contributions,
where we have  $F(v,t)=F(v)$, hence
the inner integration over $\tB$ can be done analytically and the outer
one over $v$ is done numerically.

In a general case, after swapping the integration order
\begin{equation}
\varrho^{[F]}_2(v_x,\tB_a,\tB_b;\tBl)=
\int\limits_{\tB_{\lambda} - \tB_b}^{v_x} dv 
\Bigg\{ 
  \theta_{v_x <\tB_{\lambda} - \tB_a} 
     \int\limits_{\tB_{\lambda} - v_x}^{\tB_b} d\tB\; \frac{1}{\tB + v}
 +\theta_{v_x > \tB_{\lambda} - \tB_a} 
     \int\limits_{\tB_a}^{\tB_b}               d\tB\; \frac{1}{\tB + v}
\Bigg\} F (v, \tB) 
\end{equation}
the integration is split into two parts,
first for the triangle and second for the rectangle integration area,
see Fig.~\ref{fig:sudak-rho2F} for illustration.
The above change of the integration order was also exploited
by the authors of HERWIG MC~\cite{Corcella:2000bw,Marchesini:1988cf}%
\footnote{We acknowledge private communication from Mike Seymour for on that,
  see also Chapter 5  in {\em http://hepwww.rl.ac.uk/theory/seymour/thesis/}.}.
The internal integral can be transformed using the identities:
\begin{equation}
\begin{split}
&\varrho'(\tB_b)=\partial_{\tB_b}\varrho(\tB_b;\tB_\lambda)
= \int\limits_{\tB_{\lambda}-\tB_b}^{0} dv\; \frac{1}{\tB + v}
= \theta_{\tB_b>\tB_{\lambda}}(\ln\tB_b-\ln\tB_{\lambda}),
\\&
\varrho'^{[1]}_2(v_x,\tB_a,\tB_b)
=\partial_{v_x}\varrho^{[1]}_2(v_x,\tB_a,\tB_b;\tBl)
\\&~~~~~~~~~~~
= \theta_{v_x>\tB_{\lambda}-\tB_a}
   [\varrho'(\tB_b+v_x)-\varrho'(\tB_a+v_x)]
+ \theta_{v_x<\tB_{\lambda}-\tB_a}
    \varrho'(\tB_b+v_x),
\end{split}
\end{equation}
to the following convenient form
\begin{equation}
\begin{split}
\varrho^{[F]}_2(v_x,\tB_a,\tB_b;\tBl)
&=
\int\limits_{\tB_{\lambda} - \tB_b}^{v_x} dv\;
\varrho'^{[1]}_2(v,\tB_a,\tB_b)
\int\limits_0^1 
    d\left(\frac{\ln(\tB+v)}{\varrho'^{[1]}_2(v,\tB_a,\tB_b)} \right)
F (v, \tB)
\\&
=\varrho'^{[1]}_2(v_x,\tB_a,\tB_b) 
 \int\limits_0^1 d\xi(v)
 \int\limits_0^1 d\sigma(v,\tB)\;
F (v, \tB),
\end{split}
\end{equation}
where
\begin{equation}
 \xi(v)=\frac{\varrho^{[1]}_2(v,\tB_a,\tB_b)}{\varrho^{[1]}_2(v_x,\tB_a,\tB_b)},\quad
\sigma(v,\tB)
= \frac{\ln(\tB+v)}{\varrho'^{[1]}_2(v,\tB_a,\tB_b)}.
\end{equation}
The above mapping is used in the bremsstrahlung CMC in cases (B) and (C).
In this context one also needs to perform the inverse mapping $v(\xi)$,
which requires numerical inversion of $\varrho^{[1]}_2(v,\tB_a,\tB_b)$
as a function of $v$.
Once $v$ is known, the second inverse mapping $t(\sigma,\xi)$
is easily implemented, as it can be formulated in a fully analytical form.
As already indicated, a very similar variant of the above integration order
is used in the evaluation of the non-IR and flavour-changing parts
of the Sudakov form-factor.

\subsection{Integration area in the plane of $\ln k^T$ and rapidity $\eta$}
\label{append:splane}
Finally, let us relate the phase space $v$ and $t$ used in the calculation
of the form-factor above with the Sudakov plane in $\ln k^T$ and the rapidity $\eta$.
This is done in the pictorial way in Fig.~\ref{fig:sudak0}, where
we depict a situation with three emission, underlining the trapezoid
integration domain for the Sudakov function $\Phi(t_3,t_2|x_2)$.
This is show on the right hand side of the figure,
as the  trapezoid marked by $abcd$, in the plane of $t$ and $v$.
On the left hand side of this figure, the corresponding trapezoid is seen
in the plane of $\ln k^T$ and rapidity $\eta$.
Let us remind the reader that
in the case (C) we define $v_i=\ln y_i=\ln(x_i-x_{i-1})$ and
the rapidities $\eta_i$ are related to the evolution times $t_i$ with the simple linear
transformation of eq.~(\ref{eq:dwojka}) in Sec.~\ref{sec:kinema}.


\providecommand{\href}[2]{#2}

\end{document}